\documentclass[useAMS,usenatbib]{mnras}
\usepackage{aasmacros}
\usepackage{amsmath,times}
\usepackage{graphicx,url}
\usepackage{amssymb}
\usepackage[usenames,dvipsnames]{color}
\usepackage[normalem]{ulem}
\title[Substructure and Galaxy Formation in Warm Dark Matter]{
  Substructure and galaxy formation in the Copernicus Complexio warm
  dark matter simulations} \author[S. Bose et al.]{Sownak
  Bose$^{1}$\thanks{E-mail: sownak.bose@durham.ac.uk}, Wojciech
  A. Hellwing$^{2,3}$, Carlos S. Frenk$^{1}$, Adrian Jenkins$^{1}$,
  \newauthor Mark R.  Lovell$^{4,5}$, John C. Helly$^{1}$, Baojiu
  Li$^{1}$, Violeta Gonzalez-Perez$^{2}$, and Liang Gao$^{1,6}$
  \\ $^{1}$Institute for Computational Cosmology, Durham University,
  South Road, Durham, UK, DH1 3LE \\ $^{2}$Institute of Cosmology and
  Gravitation, University of Portsmouth, Portsmouth, UK, PO1 3FX
  \\ $^{3}$Janusz Gil Institute of Astronomy, University of Zielona
  G{\' o}ra, ul. Szafrana 2, 65-516 Zielona G{\' o}ra, Poland
  \\ $^{4}$GRAPPA Institute, Universiteit van Amsterdam, Science Park
  904, 1098 XH Amsterdam, The Netherlands \\ $^{5}$Instituut-Lorentz
  for Theoretical Physics, Niels Bohrweg 2, NL-2333 CA Leiden, The
  Netherlands \\ $^{6}$National Astronomical Observatories, Chinese
  Academy of Sciences, 20A Datun Road, Chaoyang District, Beijing
  100012, China \\}

\newcommand{\Msun}{h^{-1}\,M_\odot}

\newcommand{\vmax}{$V_{\rmn{max}}$}

\def\gsim{ \lower .75ex \hbox{$\sim$} \llap{\raise .27ex \hbox{$>$}} }
\def\lsim{ \lower .75ex \hbox{$\sim$} \llap{\raise .27ex \hbox{$<$}} }

\def\jcap{JCAP}
\newcommand{\bq}{\begin{eqnarray}}
\newcommand{\eq}{\end{eqnarray}}

\def\cocow~{\textsc{coco-warm}}
\def\cococ~{\textsc{coco-cold}}

\begin{document}

    \pagerange{\pageref{firstpage}--\pageref{lastpage}} \pubyear{2016}

  \maketitle

  \label{firstpage}
  
\begin{abstract}
  We use the {\it Copernicus Complexio} (\textsc{coco}) high
  resolution $N$-body simulations to investigate differences in the
  properties of small-scale structures in the standard cold dark
  matter (CDM) model and in a model with a cutoff in the initial power
  spectrum of density fluctuations consistent with both a thermally
  produced warm dark matter (WDM) particle with a rest mass of 3.3
  keV, or a sterile neutrino with mass 7~keV and leptogenesis
  parameter $L_6=8.7$. The latter corresponds to the ``coldest'' model
  with this sterile neutrino mass compatible with the identification
  of the recently detected 3.5~keV X-ray line as resulting from
  particle decay. CDM and WDM predict very different number densities
  of subhaloes with mass $\lsim 10^9\,\Msun$ although they predict
  similar, nearly universal, normalised subhalo radial density
  distributions. Haloes and subhaloes in both models have cuspy NFW
  profiles, but WDM subhaloes below the cutoff scale in the power
  spectrum (corresponding to maximum circular velocities
  $V_{\mathrm{max}}^{z=0} \leq50~\mathrm{kms}^{-1}$) are less
  concentrated than their CDM counterparts. We make predictions for
  observable properties using the \textsc{galform} semi-analytic model
  of galaxy formation. Both models predict Milky Way satellite
  luminosity functions consistent with observations, although the WDM
  model predicts fewer very faint satellites. This model, however,
  predicts slightly more UV bright galaxies at redshift $z>7$ than
  CDM, but both are consistent with observations. Gravitational
  lensing offers the best prospect of distinguishing between the
  models.
\end{abstract}
 
 
\section{Introduction} 
\label{Intro}

Over the past three decades, advances in the scale and sophistication
of numerical simulations have led to significant progress in studies
of the non-linear phases of cosmological structure formation.
Simulations have played a major role in establishing Lambda Cold Dark
Matter ($\Lambda$CDM, hereafter just CDM) as the standard model of
cosmogony, helping link the theory to observations over a large range
of scales and epochs, from temperature fluctuations in the cosmic
microwave background (\citealt{Planck2013}) through the large-scale
distribution of galaxies
(\citealt{2dfgrs,Zehavi2002,Hawkins2003,sdss,bao_2df,bao_sdss}) to the
inner structure of dark matter haloes \citep[see][for a
  review]{Frenk_White2012}.

Although it is almost certainly the case that the dark matter consists
of non-baryonic elementary particles \citep{Planck2013}, the identity
of the particle remains a mystery.  There are hotly debated claims
that dark matter of different kinds has been indirectly detected. The
``gamma-ray excess'' observed towards the Galactic centre has been
ascribed to annihilation radiation of cold dark matter
\citep{Hooper2011}. On the other hand the 3.53~keV X-ray line detected
in the stacked spectrum of clusters (\citealt{Bulbul2014a}) and,
independently, in the Perseus cluster and Andromeda
(\citealt{Boyarsky2014a}) has been ascribed to the decay of a 7~keV
sterile neutrino (but see, e.g.
\citealt{Malyshev2014,Anderson2015,Sorensen2016} for different
interpretations). These two kinds of dark matter would produce very
similar large-scale structure but can give rise to large observable
differences on small scales.

Warm dark matter (WDM) particles have non-negligible thermal
velocities at early times which damp primordial density fluctuations
below a free streaming scale. The position of the cutoff depends on
the mass and the production mechanism of the warm particles. If they
are thermally produced, the cutoff length scales inversely with the
particle mass and, if the particle mass is in the keV range, the
cutoff corresponds roughly to the scale of dwarf galaxies. Structure
formation in such models has been extensively studied using
simulations in the past few years
(e.g. \citealt{Colin2000,Bode2001,Viel2005,Knebe2008,Schneider2012,Lovell2012,
  Maccio2013,Lovell2014,Reed2015,Colin2015,Yang2015,Bose2016,Horiuchi2016}).
The observed clumpiness of the Lyman-$\alpha$ forest sets a lower
limit to the mass of a dominant thermally produced WDM particle of
$m_{\mathrm{WDM}} \geq 3.3$ keV at $95\%$ confidence \citep{Viel2013};
this is consistent with a lower limit set by the observed abundance of
satellites in the Milky Way \citep{Kennedy2014,Lovell2015}. The lower
limit to the mass of thermal WDM was increased to $m_{\mathrm{WDM}}
\geq 4.35$ keV (95\% confidence) by \cite{Baur2016}, who repeated the
analysis of \cite{Viel2013} with an updated sample of QSO spectra from
SDSS-III. These limits, however, depend on uncertain assumptions for
thermal history for the intergalactic medium \citep{Garzilli2015}.

Warm dark matter in the form of sterile neutrinos can also form
through resonant processes in the early universe \citep{Shi1999}. In
this case, the primordial fluctuation spectrum is more complicated and
both the position and shape of the cutoff depend on the formation
mechanism. In the ``neutrino Minimal Standard Model'' ($\nu$MSM,
\citealt{Asaka2005,Boyarsky2009}) (a simple extension of the Standard
Model of particle physics) right-handed sterile neutrinos of keV mass
($M_1$) make up a triplet alongside two other neutrinos of GeV mass
($M_2$ and $M_3$). $M_1$ behaves as WDM. If $M_2$ and $M_3$ decay
before the production of the dark matter, they can create a ``lepton
asymmetry'' (i.e. an excess of leptons over antileptons). In the
presence of this asymmetry, the production of the dark matter can be
boosted by resonant production (\citealt{Shi1999}). The result is a
model with the correct abundance of dark matter that  also explains
active neutrino oscillations.  

\begin{figure}
\centering
\includegraphics[width=\columnwidth]{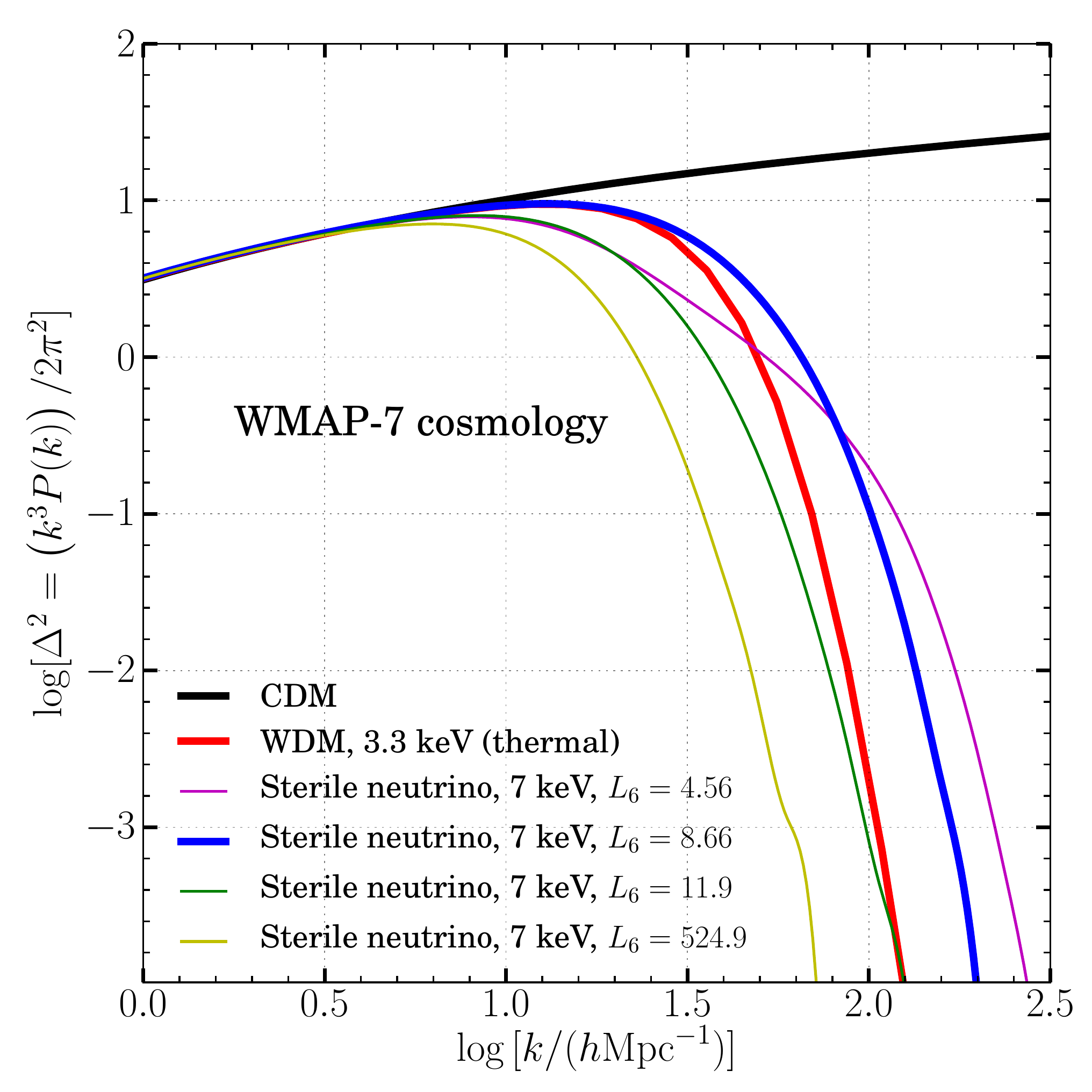}
\caption{The (dimensionless) matter power spectrum for a thermal
  $3.3$~keV WDM (red, used in \cocow~), a sterile neutrino of mass
  $m_{\nu_s} = 7$ keV and lepton asymmetry $L_6 = 8.66$ (blue) and CDM
  (black, used in \cococ~).  Power is significantly suppressed at
  small scales for both thermal WDM and the sterile neutrino.  The
  deviation from CDM occurs at almost identical scales:
  $\log\left[k(h/\rm{Mpc}^{-1})\right] \gtrsim 1.0$. The excess of
  power at high-$k$ for $L_6 = 8.66$ compared to the 3.3 keV case is
  negligible ($\lesssim 1\%$) for the scales resolved in our
  simulations; see Appendix B in \citet{Lovell2015}. The other thin
  coloured lines show power spectra for 7~keV sterile neutrinos with
  different values of $L_6$, as labelled in the legend. Figure
  reproduced from \citet{Bose2016}.}
\label{powerspec}
\end{figure}

The leptogenesis parameter, $L_6$, which determines the asymmetry,
affects the cutoff scale and shape of the power spectrum cutoff in a
non-trivial way illustrated in Fig.~\ref{powerspec}. The CDM power
spectrum is shown as a thick black line and sterile neutrino models
with particle mass of 7~keV and $L_6$ ranging from 4.6 to 525 are
shown by colour lines. All these power spectra are essentially
identical on scales larger than $\log\left[k/(h\rm{Mpc}^{-1})\right]
\sim 0.5$, but differ on smaller scales both in the shape and location
of the cutoff which, furthermore, does not vary monotonically with
$L_6$.  The power spectrum of a thermal 3.3~keV WDM model (the
limiting mass consistent with the Lyman-$\alpha$ forest constraint of
\citealt{Viel2013}) is also plotted. This has a similar cutoff scale
as the $L_6 = 8.66$ model, which is the ``coldest'' possible sterile
neutrino model with a 7~keV particle mass.

In this paper we use the {\it Copernicus Complexio}
(\textsc{coco-warm}) high resolution $N$-body simulation to investigate
the properties of subhaloes in a WDM model. This simulation was run
prior to the discovery of the 3.5~keV line and its initial power
spectrum was chosen to correspond to a thermal 3.3~keV WDM model. This
turns out to have been a fortuitous choice since this power spectrum
is very similar to that of the coldest 7~keV sterile neutrino, so
constraints on this model can be readily extended to all sterile
neutrino models with a 7~keV particle mass. The formation times, mass
functions, spins, shapes, mass profiles and concentrations of haloes in
this simulation were presented in \cite{Bose2016}.  Here we focus on
the properties of halo substructures. 

The \textsc{coco} simulations are amongst the highest resolution WDM
$N$-body simulations of a cosmological volume performed to date (see
Section~\ref{SimsOverall}.  Previous simulations at higher mass
resolution have focussed on properties of individual haloes
\citep[e.g.][]{Lovell2014,Colin2015}. Other WDM simulations of
comparable mass resolution to ours \citep[e.g.][]{Schneider2013}
followed smaller volumes. The advantage of the relatively high mass
resolution and large volume of \textsc{coco} is that it provides a
large statistical sample of well-resolved dark matter haloes spanning
nearly seven decades in mass. In particular, resolving the halo mass
function down to $ \sim 10^7-10^8\,h^{-1}\,M_\odot$, as \textsc{coco}
does, is a crucial input to studies that attempt to distinguish
amongst different types of dark matter using strong gravitational
lensing \citep{Vegetti2009,RanLi2016}.

Our simulations are numerically converged down to a halo peak circular
velocity of $V_{\rm{max}}, \geq 10\,\rm{kms}^{-1}$, thus allowing
statistically meaningful studies of the satellites of the Milky
Way. Furthermore, the high resolution of our simulations makes it
possible to construct accurate merger trees of even such small haloes
and, as a result, we can calculate their observable properties, using
the Durham semi-analytical galaxy formation model, \textsc{galform}
\citep{Cole2000,Lacey2015}, a flexible and effective method to
implement the best current understanding of galaxy formation physics
into an $N$-body simulation.

The layout of this paper is as follows. In Section~\ref{SimsOverall}
we introduce the \textsc{coco} simulation set, which includes both the
WDM model of interest here and its CDM counterpart.  In
Section~\ref{substructureDM} we investigate the main properties of
subhaloes: their population statistics, distribution and internal
structure. In Section~\ref{galaxygen} we describe the \textsc{galform}
model and the modifications required for the WDM case, and compare to
predictions for the CDM case.  Finally, we summarise our results in
Section~\ref{conclusion}.

\section{The Simulations}
\label{SimsOverall}

\begin{table*}
\centering
\vspace{10pt}
\begin{tabular}{cccccccc}
\hline \hline \\
Simulation & Cube length & $m_{\rm{WDM}}$ & $N_p$ & $V_{\rm{hr}}$ & 
$m_{p,\rm{hr}}$ &  $N_{p,\rm{hr}}$ & $\epsilon_{\rm{hr}}$\\ \\
\hline \hline \\
\textsc{color} & $70.4\,h^{-1}\,\rm{Mpc}$ & 3.3 keV & $4,251,528,000$ & 
$70.4^3\,h^{-3}\,\rm{Mpc}^3$ 
& $6.196 \times 10^6\,h^{-1}\,M_\odot$ & $4,251,528,000$ & $1\,h^{-1}\,\rm{kpc}$ 
\\  
\textsc{coco} & $-$ & 3.3 keV & $13,384,245,248$ & $\sim 2.2 \times 
10^4\,h^{-3}\,\rm{Mpc}^3$ 
& $1.135 \times 10^5\,h^{-1}\,M_\odot$ & $12,876,807,168$ & 
$230\,h^{-1}\,\rm{pc}$ \\ 

\\ \hline \hline
\end{tabular}
\caption{Simulation parameters assumed in \textsc{coco} and the parent
  simulation, \textsc{color}. Here, $m_{\mathrm{WDM}}$ is the mass of
  the thermal relic warm dark matter particle; $N_p$ is the total
  number of particles (of all types) used in the simulation;
  $V_{\mathrm{hr}}$ is the approximate volume of the high-resolution
  region at $z=0$; $m_{p, \mathrm{hr}}$ is the mass of an individual
  high-resolution dark matter particle; $N_{p, \mathrm{hr}}$ is the
  total number of particles of this species; and
  $\epsilon_{\mathrm{hr}}$ is the softening length.  The parameter
  $m_{\mathrm{WDM}}$ is only relevant for \cocow~.}
\label{tableparams}
\end{table*}

\subsection{Simulation details}

The $N$-body simulations used in this work, \textsc{coco}, were
introduced by \cite{Hellwing2016} and \cite{Bose2016}, as part of the
{\em Virgo Consortium} programme. In short, \textsc{coco} is a set of
cosmological zoom-in simulations that follow about 12 billion high
resolution dark matter particles, each of mass $m_p = 1.135 \times
10^5 \Msun$. The resimulation region was extracted from the
$(70.4\,h^{-1}~{\rm Mpc})^3$ parent volume, {\em Copernicus complexio
  Low Resolution} (\textsc{color}). \textsc{color} and \textsc{coco}
assume cosmological parameters obtained from WMAP-7: $\Omega_m =
0.272, \Omega_\Lambda = 0.728, h = 0.704, n_s = 0.967$ and $\sigma_8 =
0.81$. The simulations were performed using \textsc{gadget-3}, an
updated version of the publicly available \textsc{gadget-2} code
\citep{GADGET,GADGET2,Springel2008}.  Substructures in the simulation
were identified using the \textsc{subfind} algorithm
\citep{Springel2001}. For a comprehensive description of the initial
conditions and choice for the zoom-in region, we refer the reader to
\cite{Hellwing2016} and \cite{Bose2016}.

\begin{figure*}
\centering
\includegraphics[width=\textwidth,height=0.5\textheight]{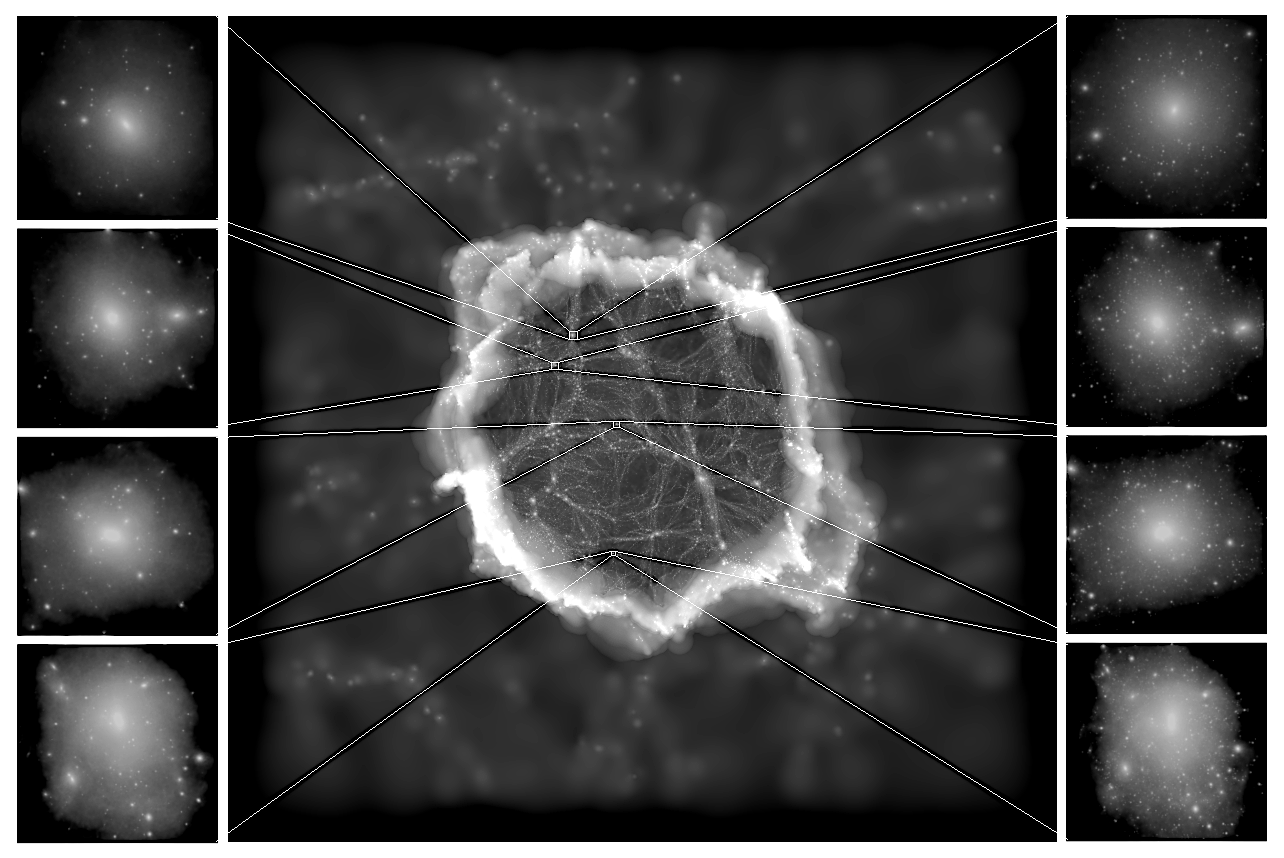}
\caption{Projected density map in a slice of dimensions $\left(70.4
  \times 70.4 \times 1.5\right)\,h^{-1}\,\rm{Mpc}$ centred on the
  \textsc{coco} high resolution region at $z=0$. The intensity of the
  image scales with the number density of particles in the region. The
  side panels show zooms of a sample of haloes identified at $z=0$,
  matched between \cocow~ (left) and \cococ~ (right)}
\label{image}
\end{figure*}

\textsc{coco} consists of two simulations with initial fluctuation
power spectra corresponding to CDM (\cococ~) and to the 3.3~keV
thermal relic WDM (\cocow~); the initial fluctuation field had the
same phases in both cases. \cocow~ has a power spectrum with a
sharp cutoff at small scales, which is approximated by the fitting
formula:
\bq \label{transfer}
T(k) = \left( 1+\left(\alpha k \right)^{2\nu}\right)^{-5/\nu},
\eq
\citep{Bode2001}, where $T(k)$ is the transfer function relating the
power spectra for CDM and WDM, $\nu = 1.12$ is a constant, and
$\alpha$ is determined by the mass of the particle:
\bq \label{alphaMass}
\alpha = 0.049 \left[ \frac{m_{\mathrm{{\tiny WDM}}}}{\mathrm{keV}} 
\right]^{-1.11} \left[ \frac{\Omega_{\mathrm{{\tiny WDM}}}}{0.25} 
\right]^{0.11}\left[\frac{h}{0.7}\right]^{1.22}~h^{-1}~\mathrm{Mpc}.
\eq
\citep{Viel2005}.  The CDM and WDM power spectra are then related by:
\bq
P_{\mathrm{WDM}} (k) = T^2(k) P_{\mathrm{CDM}} (k).
\eq
The power spectra used in the \textsc{coco} simulations are shown as
thick lines in Fig.~\ref{powerspec} and are discussed in
Section~\ref{Intro}.  The main simulation parameters are listed in
Table~\ref{tableparams}. A projected density map of the \textsc{coco}
volume at $z=0$ is shown in Fig.~\ref{image}.

\subsection{Subhalo mass definitions}

The mass of a halo, $M_{\Delta}$, is defined as the mass within
$r_{\Delta}$, the radius within which the average density is $\Delta$
times the critical density of the Universe at the redshift at which
the halo is identified. Usually $\Delta = 200$ is used to define the
spherical overdensity within the virialised region of the halo, but we
will also use $\Delta = 50$ in order to compare with the results of
the \textsc{aquarius} project simulations
\citep{Springel2008}.\footnote{Note that in the \cite{Springel2008},
  $r_{50}$ is the radius as which the mean density is $200 \Omega_m$.
  In \textsc{coco}, this would correspond to roughly $r_{54}$.} For
the mass of a subhalo, $M_{\mathrm{sub}}$, we take the mass identified
by \textsc{subfind} as the mass that is gravitationally bound to the
subhalo.

\subsection{Identification and removal of numerical artefacts}
\label{spurious}

Since our analysis is primarily concerned with the properties of dark
matter subhaloes and the galaxies that form in them, it is important
to ensure that the merger trees have been pruned of the artificial
structures that form from evolution from a power spectrum with a
resolved cutoff.  Spurious structures along filaments were apparent in
the first WDM simulations \citep{Bode2001} but were only subsequently
recognised as an effect of particle discreteness by \cite{Wang2007}.
A technique to eliminate spurious objects in post-processing was
developed by \citep{Lovell2014}, while \cite{Angulo2013} and
\cite{Hobbs2016} have proposed modifications to the $N$-body simulation method
itself to prevent the formation of such objects in the first place.
  
\cite{Wang2007} found that a large fraction of the spurious haloes
can be removed by rejecting objects with mass below a resolution-dependent
threshold:
\bq
M_{\mathrm{lim}} = 10.1 \bar{\rho}~ d~ k_{\mathrm{peak}}^{-2}\;,
\eq
where $\bar{\rho}$ is the mean matter density in the universe, $d$ is
the mean interparticle separation and $k_{\mathrm{peak}}$ is the
spatial frequency at which the dimensionless input power spectrum,
$\Delta^2(k)$, has its maximum. \cite{Lovell2014} found that particles
in spurious haloes in WDM originate from Lagrangian patches that are
much flatter than the corresponding Lagrangian patches in CDM.

\begin{figure}
\centering
\includegraphics[width=\columnwidth]{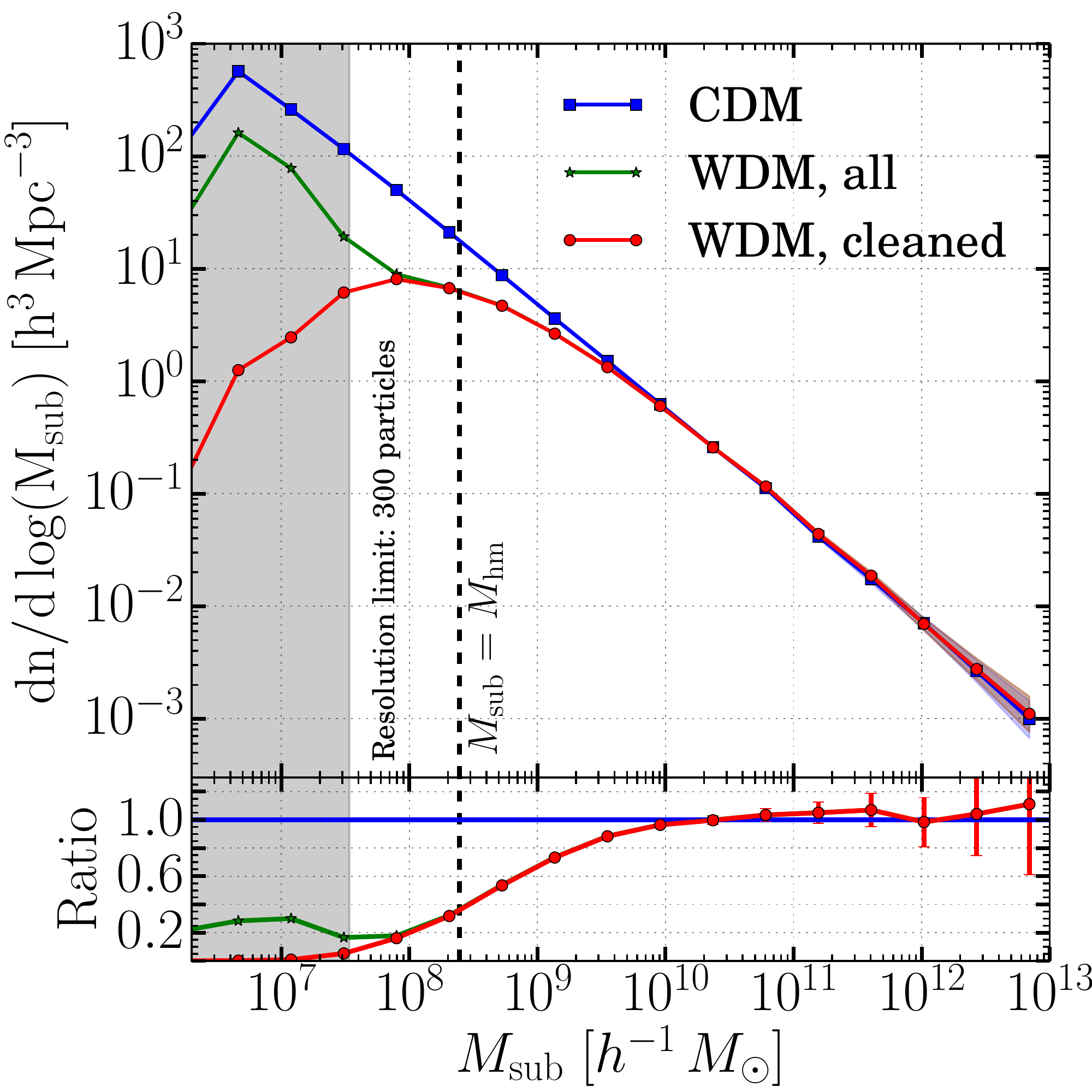}
\caption{Upper panel: the number density of subhaloes as a function of
  subhalo mass, $M_{\mathrm{sub}}$, for \cococ~ (blue), \cocow~ with
  all objects included (green), and \cocow~ with spurious
  structures removed (red). The shaded region around each curve
  represents the Poisson uncertainty in the number counts
  in that bin. The vertical black dashed line marks the half-mode mass,
  $M_{\rm{hm}}$, for the 3.3 keV thermal relic. The grey shaded region 
  demarcates the resolution limit
  of our simulations, set at 300 particles, which was determined by
  requiring convergence of the mass function compared with the
  lower-resolution version of \cococ~, \textsc{color-cold}. Lower
  panel: the ratio of the two \cocow~ mass functions to 
  the \cococ~ mass function.} 
\label{submassfunc}
\end{figure}

\cite{Lovell2014} devised the following procedure for identifying
spurious haloes. Defining $M_{\mathrm{max}}$ as the maximum mass
attained by a halo during its evolution, and $s_{\mathrm{half-max}}$
as the sphericity ($c/a$, where $c$ is the minor and $a$ the major
axis of a uniform density ellipsoid with the same inertia tensor) of
the halo particles (at high redshift) when it attains half of its 
maximum mass, we: (1)~discard all (sub)haloes with 
$s_{\mathrm{half-max}} < s_{\rm cut}$, irrespective
of mass, and (2)~for those that pass (1), remove (sub)haloes with
$M_{\mathrm{max}} < 0.5 M_{\mathrm{lim}}$.  The threshold sphericity
in step (1) is chosen such that 99\% of CDM haloes are rounder than
$s_{\rm cut}$. This threshold sphericity needs to be calculated for
haloes identified at each redshift; at $z=0$, $s_{\rm cut}=0.165$.  The
factor of 0.5 in step (2) was obtained by comparing different
resolution levels of warm dark versions of the \textsc{aquarius}
simulations \citep[see][for details]{Lovell2014}.  We apply this
procedure to (sub)haloes in every snapshot in our simulation; branches
in the merger tree that contain a spurious object are pruned from the
tree.

\section{Dark Matter Substructure}
\label{substructureDM}

In this section we study the dark matter substructure in the \cococ~
and \cocow~ simulations, quantifying their abundance, distribution and
internal structure. The general trend we find is that the largest
subhaloes in \cocow~ and \cococ~ are indistinguishable but differences
become increasingly significant below $\sim 5 \times 10^{9}\,h^{-1}\,M_\odot$.

\begin{figure}
\centering
\includegraphics[width=\columnwidth]{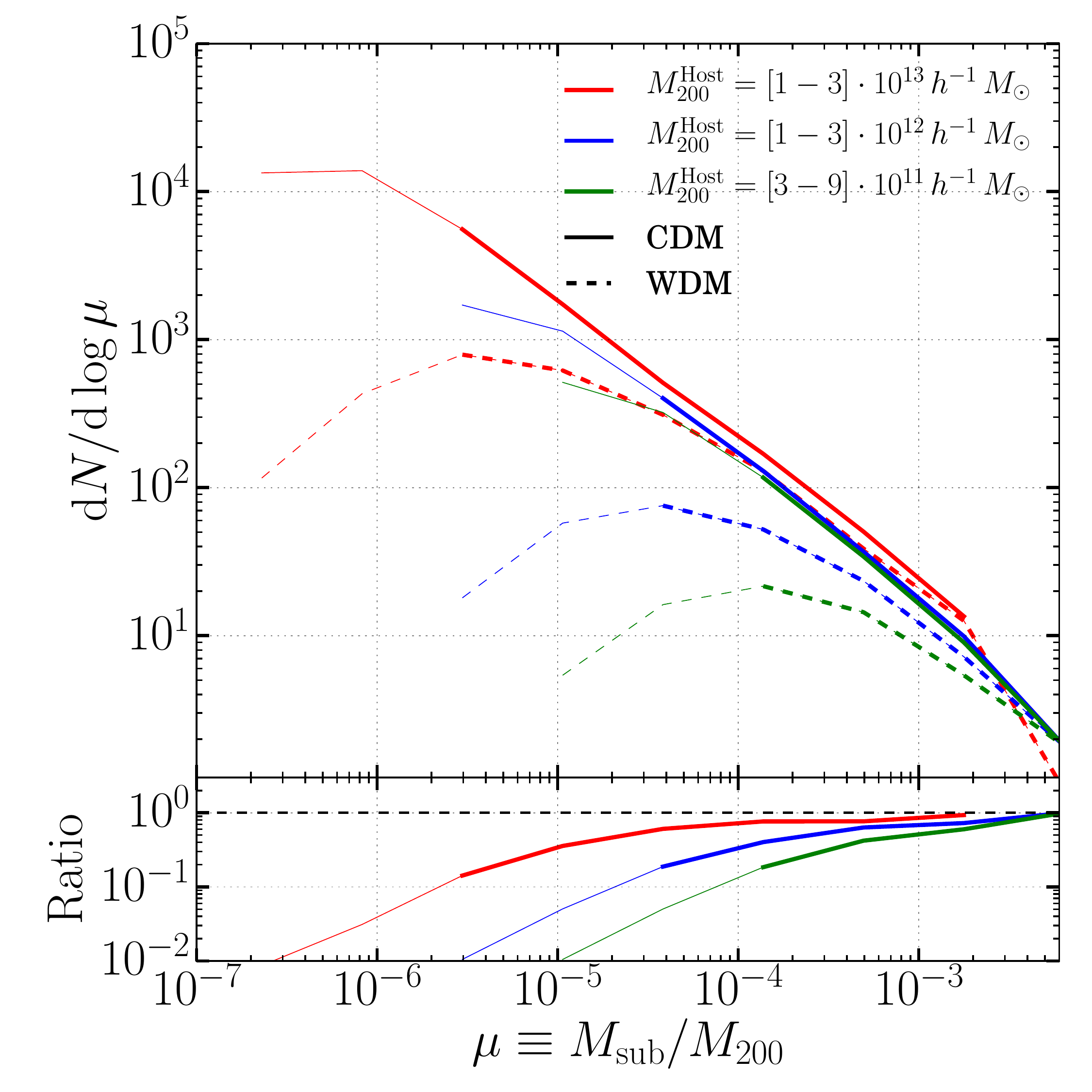}
\caption{Upper panel: the stacked differential subhalo mass function as a
  function parent halo mass, expressed in units of $M_{\mathrm{sub}} /
  M_{200}$. The CDM case is shown with solid lines and the WDM
  case with dashed lines. The different colours correspond to
  different host halo mass ranges as indicated in the legend. The
  lines become thinner when a given subhalo has fewer than 300 particles
  i.e., when $\mu \times M_{200,\rm{mid}}^{\rm{host}} > 300 m_p$,
  where $M_{200,\rm{mid}}^{\rm{host}}$ is the centre of the host halo mass
  bin, and $m_p$ is the high resolution particle mass.  Lower panel: ratio
 of the differential subhalo mass functions in WDM to those in CDM.} 
\label{parentmsub}
\end{figure}

\begin{figure}
\centering
\includegraphics[width=\columnwidth]{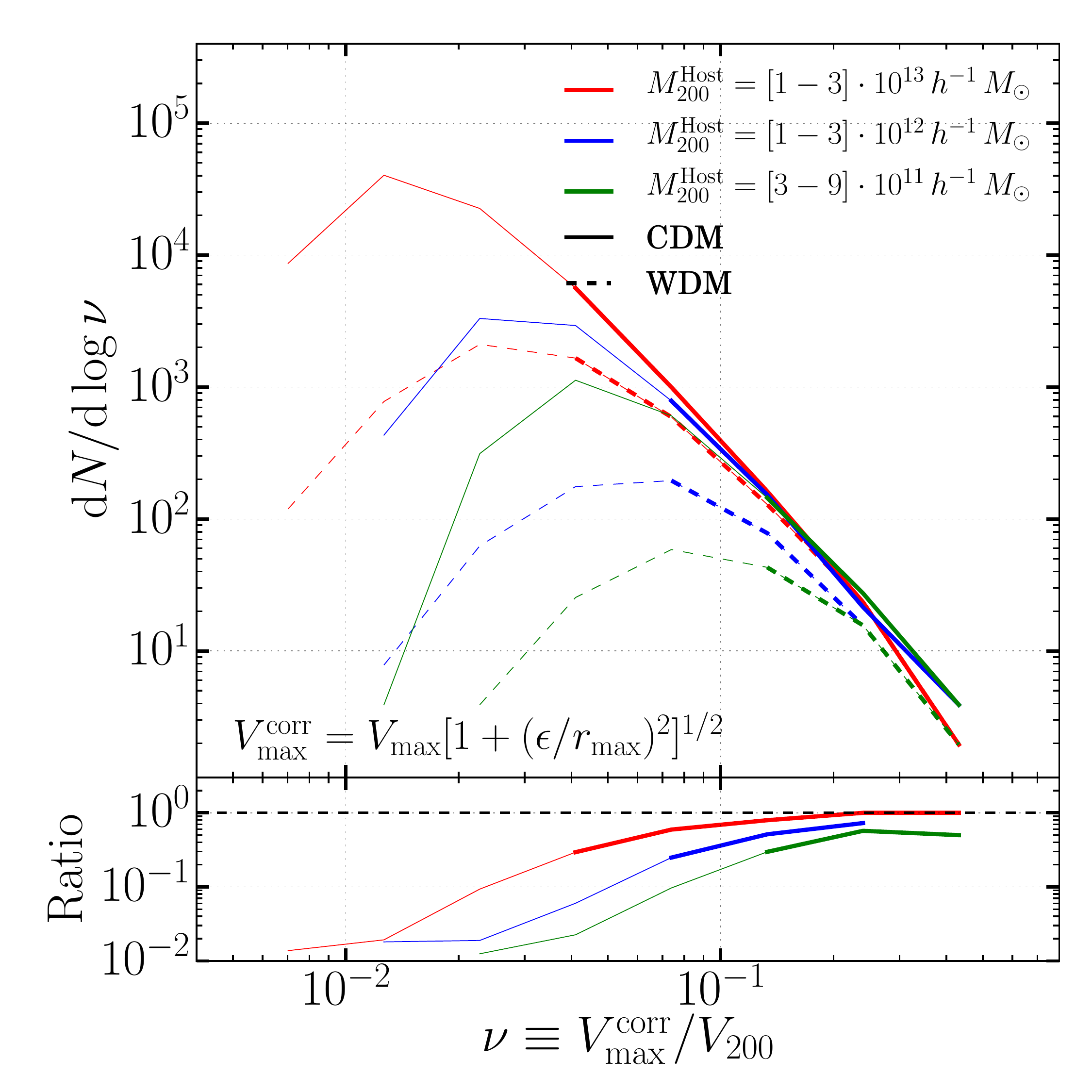}
\caption{As Fig.~\ref{parentmsub}, but with subhalo abundance
  expressed as a function of $V_{\mathrm{max}}^{\mathrm{corr}} /
  V_{200}$, where $V_{\mathrm{max}}^{\mathrm{corr}}$ is the maximum
  circular velocity, $V_{\mathrm{max}}$, corrected for the effects of
  gravitational softening as indicated in the legend (see main
  text). The lines become thinner when $V_{\mathrm{max}} <
  10\,\mathrm{kms}^{-1}$, which is the circular velocity to which the
  simulations are complete.}
\label{parentvmax}
\end{figure}

\begin{figure*}
\centering
\includegraphics[height=0.35\textheight, 
width=0.8\textwidth]{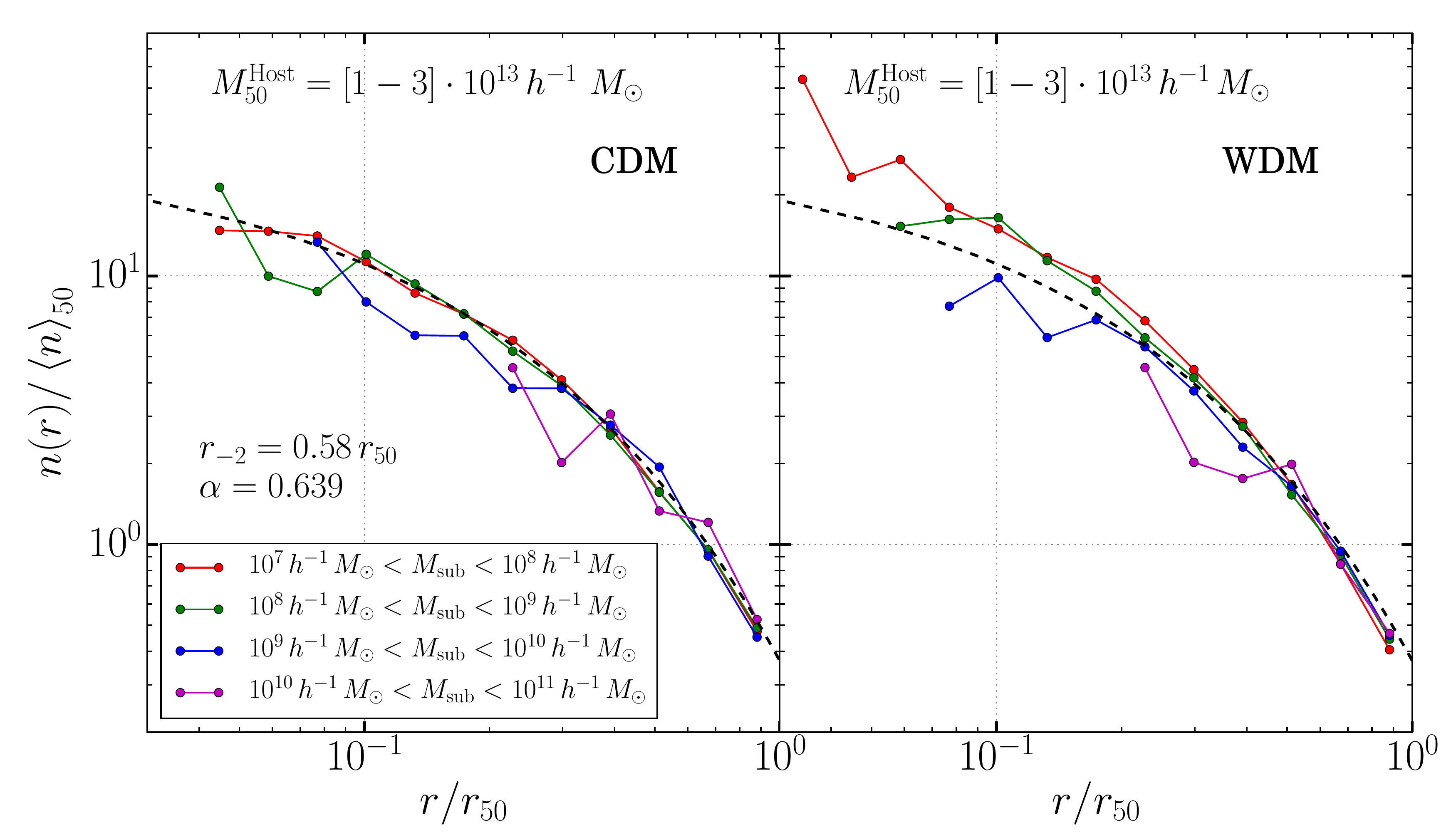}
\caption{Stacked radial number density profiles of subhaloes, $n(r)$,
  in different mass ranges (different colours), normalised to the mean
  number density in that mass range within $r_{50}$
  ($\left<n\right>_{50}$). The profiles are plotted as a function of
  the distance from the host halo centre (with mass
  $M_{50}^{\mathrm{Host}} = [1-3] \cdot 10^{13}\,h^{-1}\,M_\odot$). Left: CDM;
  right: WDM. The dashed black line shows the Einasto profile fit to
  the \cococ~ profiles, with the fit parameters $r_{-2}$ and $\alpha$
  quoted in the plot. Only subhaloes with more than 300 particles are
  shown.}
\label{subradialprofile}
\end{figure*} 

\begin{figure*}
\centering
\includegraphics[height=0.6\textheight, 
width=0.8\textwidth]{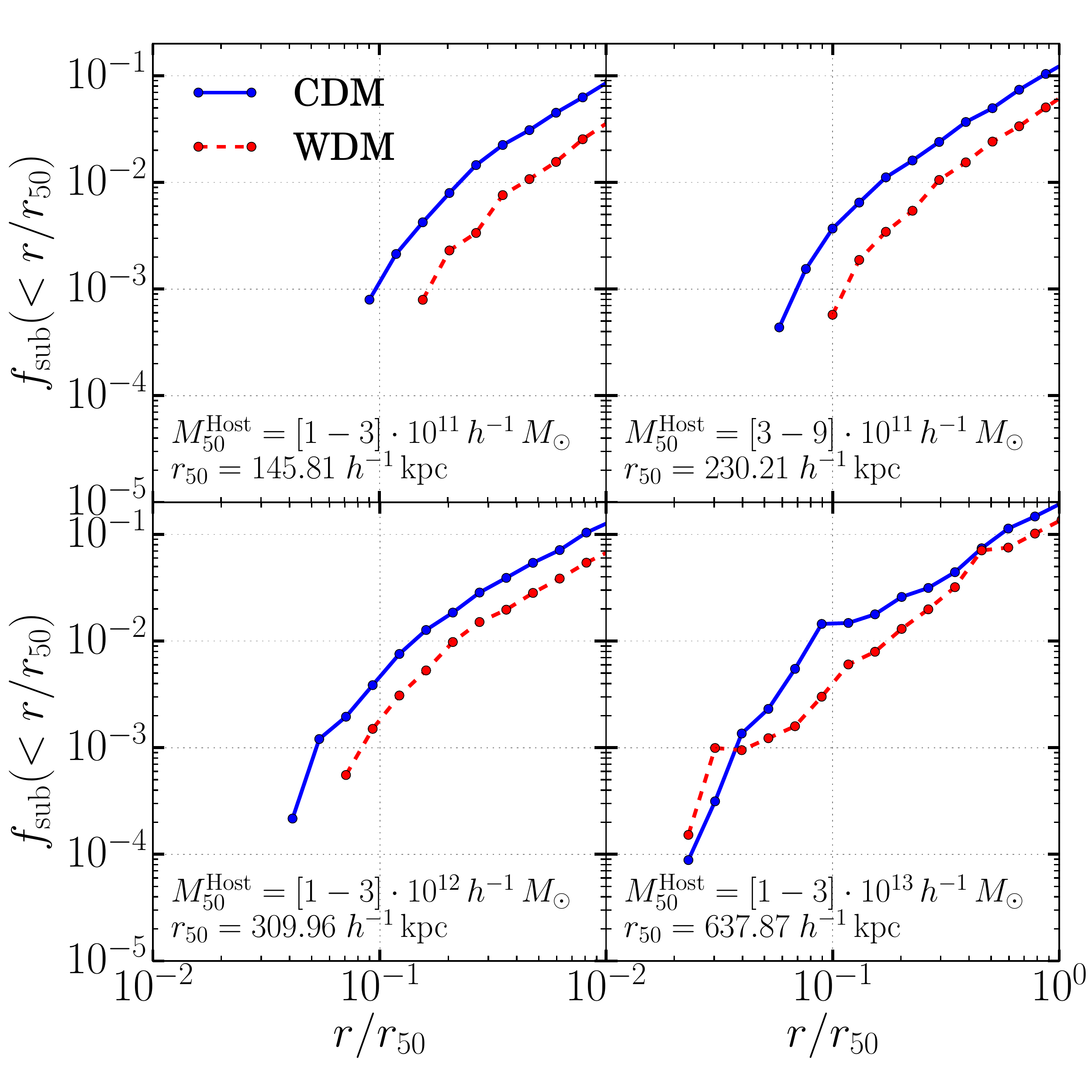}
\caption{The mass fraction in substructures as a function of
  dimensionless radial distance from the halo centre, $r/r_{50}$, for
  \cococ~ (solid blue) and \cocow~ (dashed red) at $z=0$. The four
  different panels show results for stacks of host haloes of different
  mass as indicated in the legend. Only subhaloes with more than 300
  particles are included. The value of $r_{50}$ quoted in each panel
  is the mean over all haloes in each (\cococ~) mass bin (the values
  are similar for \cocow~).}
\label{fracmass}
\end{figure*}

\begin{figure}
\centering
\includegraphics[width=\columnwidth]{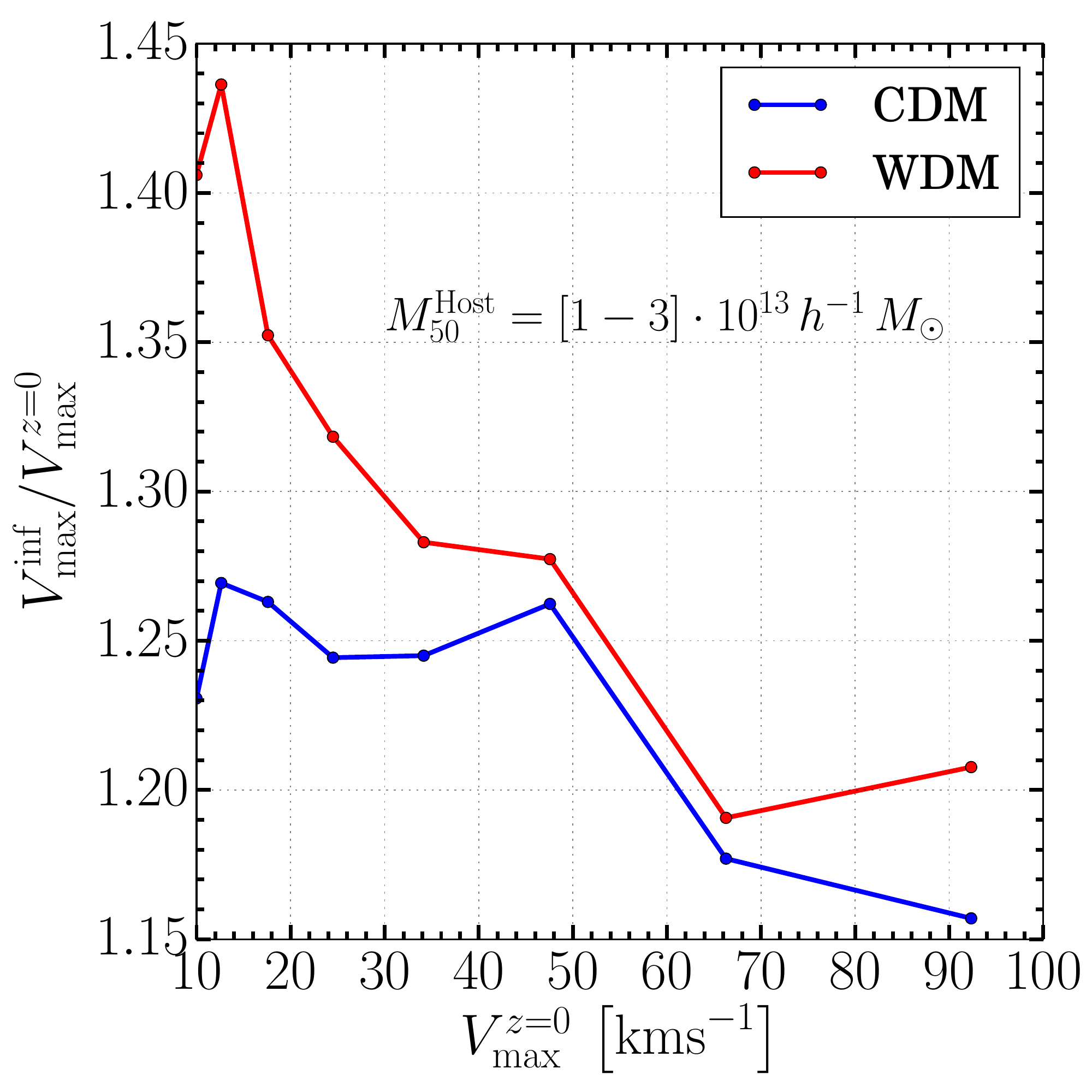}
\caption{Ratio of the infall ($V_{\mathrm{max}}^{\mathrm{inf}}$) to
  present day ($V_{\mathrm{max}}^{z=0}$) circular velocity, as a
  function of the present-day circular velocity. The results shown are
  for 6 stacked host haloes in the mass range $M_{50}^{\mathrm{Host}}
  = [1-3] \cdot 10^{13}\,h^{-1}\,M_\odot$, using all subhaloes with
  more than 300 particles, located within $r_{50}$ of the host centre
  at $z=0$. The results for \cococ~ are shown in blue and for \cocow~
  in red.}
\label{vmaxstrip}
\end{figure}

\begin{figure*}
\centering
\includegraphics[width=\textwidth]{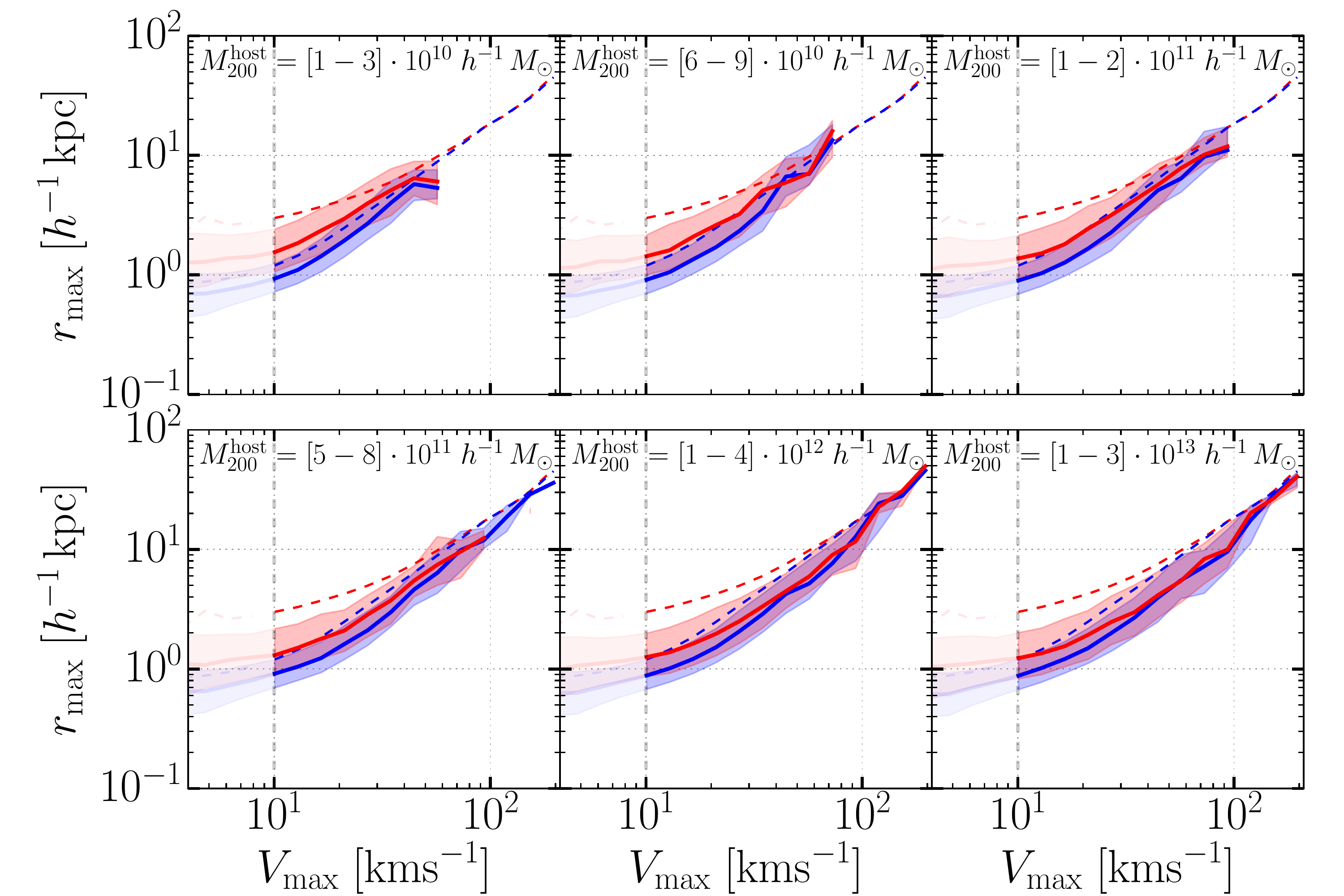}
\caption{The subhalo $r_{\mathrm{max}}$-$V_{\mathrm{max}}$ relation in
  bins of parent halo mass (different panels) for \cococ~ (blue) and
  \cocow~ (red). Each panel shows results from stacking all host
  haloes within the given mass bin. The solid line in each case shows
  the median relation in the stack, whereas the shaded regions
  correspond to the $16^{\rm{th}}$ and $84^{\rm{th}}$ percentiles.
  The dashed lines show the median relation for ``field'' haloes in
  each case.  The plots are made translucent for $V_{\mathrm{max}} <
  10~\mathrm{kms}^{-1}$, below which resolution effects become
  increasingly important (see Appendix A in \citealt{Hellwing2016}).}
\label{vrmax}
\end{figure*}

\subsection{The abundance of subhaloes}
\label{subabundance}

Fig.~\ref{submassfunc} shows the present-day differential mass function of
subhaloes, $\rm{d}n / \rm{d}\log (M_{\rm{sub}})$, as a function of mass,
$M_{\mathrm{sub}}$, in \cococ~ (blue) and \cocow~ before (green) and
after (red) the removal of artefacts. The lower panel shows the ratio
of abundances in \cocow~ relative to \cococ~. The mass function, 
$\rm{d}n / \rm{d}\log (M_{\rm{sub}})$, is normalised by noting that the 
irregular volume of the high resolution region has a mean density roughly equal 
to the mean matter density in the Universe. Combining this with the total 
mass represented by high resolution particles, we can estimate the 
volume of the high resolution region.

For $M_{\mathrm{sub}} > 10^{10}\,h^{-1}\,M_\odot$ the three mass functions
agree very well. These haloes have masses well above the free
streaming scale and no spurious objects form on these scales. Below
$M_{\mathrm{sub}}~\sim 5~\times~10^{9}~h^{-1}~M_\odot$, the \cocow~ mass
function begins to peel off from \cococ~ and by $\sim 3 \times 10^8\,h^{-1}\,
M_\odot$ it is suppressed by a factor of two. This mass is close to
the ``half-mode mass'', $M_{\mathrm{hm}}$, defined as the mass
associated with the wavenumber, $k_{\rm hm}$, at which the transfer
function in Eq.~\ref{transfer} falls to half the CDM value:
\bq
k_{\mathrm{hm}} = \frac{1}{\alpha} \left( 2^{\nu/5} -1 \right)^{1/2\nu}\; = 
\frac{2\pi}{\lambda_{\rm{hm}}} \approx 34~h\,\rm{Mpc}^{-1}.
\eq
The half-mode mass (linearly extrapolated to $z=0$) is then:
\bq
M_{\mathrm{hm}} = \frac{4\pi}{3} \bar{\rho} \left( 
\frac{\lambda_{\mathrm{hm}}}{2} \right)^{3} \approx 2.5 \times 
10^8\,h^{-1}\,M_\odot \;.
\eq
Fig.~\ref{submassfunc} shows that the abundance of subhaloes in
\cocow~ is suppressed by a factor of three at
$M_{\mathrm{hm}}$. Spurious subhaloes begin to dominate the mass
function at a mass an order of magnitude below $M_{\mathrm{hm}}$.
Before that happens, and still well above the resolution limit, at
$M_{\mathrm{sub}} \sim 10^8\,h^{-1}\,M_\odot$, the ``cleaned'' \cocow~ mass
function (i.e. after subtraction of spurious objects) is already a
factor of 5 below the CDM case and shows a sharp turnover. The lower
panel in the figure shows these trends more clearly. Removal of the
spurious subhaloes is clearly important to obtain an accurate
prediction for the abundance of low-mass galaxies in WDM models.

The statistics in {\sc coco} are good enough to allow the subhalo mass
function to be calculated for different parent (host) halo masses.
The result is shown in Fig.~\ref{parentmsub}, which gives the
(stacked) differential mass functions of subhaloes as a function of
the relative mass, $\mu \equiv M_{\mathrm{sub}}/M_{200}$ (i.e., the
subhalo mass in units of the parent halo mass), in three bins of host
halo mass. The \cococ~ functions are shown with solid lines and the
\cocow~ ones with dashed lines. In both cases, the lines become
thinner for subhaloes with fewer than 300 particles. The lower panel
of Fig.~\ref{parentmsub} shows the ratio of the differential subhalo
mass functions in \cocow~ to those in \cococ~.

The solid lines in the upper panel of Fig.~\ref{parentmsub} illustrate
the invariance of the CDM subhalo mass function, when expressed in
terms of $\mu$, previously seen by \cite{Springel2008}, \cite{Gao2012}
and \cite{Cautun2014}. The relation is well described by a nearly
universal power law (the turnover in the mass function towards low
masses is due to incompleteness caused by the resolution of the
simulations.)  The scale invariance is broken in the case of \cocow~,
where the mass function deviates from a power law at larger values of
$\mu$ for smaller host haloes.  This can be understood from the fact
that, when expressed in units of the host halo mass, the cutoff scale
(or, equivalently, $M_{\mathrm{hm}}$) is reached earlier in lower host
masses.  The abundance of subhaloes is only slightly affected for a
host of mass $M_{200} = 10^{13}\,h^{-1}\,M_\odot$, but is strongly
suppressed for $M_{200} = 10^{11}\,h^{-1}\,M_\odot$ (for which $\mu =
10^{-3}$ corresponds to $M_{\mathrm{sub}} = 10^{8}\,h^{-1}\,M_\odot$).

Given the ambiguity in the definition of subhalo mass, an alternative
property used to count bound substructures is in terms its value of
\vmax, defined as the maximum of the circular velocity
curve. Furthermore, this quantity is measurable for many real
satellites (where the rotation curve of the satellite can be measured)
so it provides a better way than the mass to compare the simulations
to observations. The upper panel of Fig.~\ref{parentvmax} shows the
``\vmax~ function,'' that is the number of subhaloes as a function of
$\nu \equiv V_{\mathrm{max}}/V_{200}$, where $V_{200}$ is the circular
velocity of the parent halo at $r_{200}$. \cite{Springel2008} found
that the convergence of the $V_{\mathrm{max}}$ function improves
markedly when \vmax~is corrected for the effects of gravitational
softening:
\bq
V_{\mathrm{max}}^{\mathrm{corr}} = V_{\mathrm{max}} \left[1 + \left( \epsilon / 
r_{\mathrm{max}}\right)^2 \right]^{1/2}.
\eq
This correction is important for subhaloes whose $r_{\mathrm{max}}$
(the radius at which \vmax~occurs) is not much larger than the
gravitational softening, $\epsilon$. The gravitational softening
adopted in \textsc{coco} ($\epsilon = 230~h^{-1}\,\mathrm{pc}$) is
quite small and we have checked that the correction does not have a
significant impact in our results. For CDM, the scale invariance of
the subhalo abundance expressed in terms \vmax~is much clearer than
when the abundance is expressed in terms of mass, as may be seen by
comparing Figs.~\ref{parentmsub} and~\ref{parentvmax}, confirming the
earlier results of \cite{Moore1999,Kravtsov2004,Zheng2005,
  Springel2008,Weinberg2008,Klypin2011,Wang2012,Cautun2014}

It is clear from Figs.~\ref{parentmsub} and~\ref{parentvmax} that,
when expressed in dimensionless units such as $\mu$ or $\nu$, the
subhalo abundance in CDM is close to universal, independent of parent
halo mass. In WDM the cutoff in the power spectrum breaks this
approximately self-similar behaviour and the subhalo abundance is no
longer a universal function.

\subsection{Radial distribution}
\label{subrad}

Perhaps surprisingly, \cite{Springel2008} found that the normalised
radial number density distribution of subhaloes is approximately
independent of subhalo mass \cite[see
  also][]{Ludlow2009,Hellwing2016}. \cite{Han2016} has provided a
simple analytical model that explains this feature, as well as the
shape of the subhalo mass function in CDM, as resulting from tidal
stripping. The subhalo radial distributions in \textsc{coco} are shown
in Fig.~\ref{subradialprofile}, which gives the radial number density
of subhaloes in different mass ranges, normalised by the mean number
density of subhaloes within $r_{50}$ at $z=0$. The distributions are
averaged over 6 parent haloes with mass in the range $1~\times~
10^{13}\,h^{-1}\,M_\odot~<~M_{50}^{\mathrm{Host}}~<~3 ~\times~
10^{13}\,h^{-1}\,M_\odot$, which are the best resolved in the
simulation.  The radial positions of the subhaloes are given in units
of $r_{50}$. Only subhaloes resolved with more than 300 particles are
included.

The dashed black lines in Fig.~\ref{subradialprofile} give a fit to
the CDM subhalo number density profiles using the Einasto profile
(\citealt{Einasto1965,Navarro2004}):
\bq
\ln \left( \frac{n}{n_{-2}}  \right) = -\frac{2}{\alpha} \left[  \left( 
\frac{r}{r_{-2}}  \right)^\alpha -1   \right]\;,
\eq
where $n_{-2}$ is the characteristic number density at the scale
radius $r = r_{-2}$.  The values of $r_{-2}$ and shape parameter,
$\alpha$, given in the legend. The fit is to \cococ~ profile and the
same curve is reproduced in the \cocow~ panel.

The fit to the CDM subhalo profile also provides an excellent fit to
the WDM profile, particularly at large radii. There are, however,
differences of detail.  The distribution of the more massive
($M_{\mathrm{sub}} > 10^9\,h^{-1}\,M_\odot$) subhaloes beyond
$r>0.2r_{50}$ is very similar in \cococ~ and \cocow~. This regime is
unaffected by the free streaming cutoff in \cocow~.  Differences in
the radial distribution of these more massive subhaloes can be
attributed to small statistics: only six $\sim
10^{13}\,h^{-1}\,M_\odot$ haloes contribute to the average shown in
Fig.~\ref{subradialprofile}. The profiles of the less massive
subhaloes ($ M_{\mathrm{sub}} < 10^9\,h^{-1}\,M_\odot$) in WDM are
somewhat steeper towards the centre than those in CDM. These subhaloes
have masses below the cutoff scale, $M_{\mathrm{hm}}$, and their
properties are affected by the cutoff. In particular, they form later
than their CDM counterparts of the same mass today and, as a result,
they have lower concentrations. These subhaloes experience more mass
loss from tidal stripping after infall.

The approximate agreement of the subhalo radial distributions
in \cococ~ and \cocow~ as well as the differences of detail are
consistent with the analytic model proposed by \cite{Han2016}.
In this model, the $z=0$ radial number density of subhaloes, $n$,
with mass, $m$, at distance, $R$, from the host halo centre is given
by:
\bq \label{haneq}
\frac{{\rm d}n(m,R)}{{\rm d}\ln m} \propto m^{-\alpha} R^{\gamma}\rho(R)\;,
\eq
where $\alpha$ is the slope of the subhalo mass function evaluated at
$m$, $\rho(R)$ is the density profile of the host dark matter halo,
$\gamma = \alpha\beta$, and $\beta \sim 1$ for an NFW density profile.
The subhalo number density profile is suppressed relative to the host
density profile by the factor $R^\gamma$.  In \cococ~, the subhalo
mass function follows a single power law but, in \cocow~, it has the
same slope as in \cococ~ for $M_{\rm{sub}} \geq
10^{10}\,h^{-1}\,M_\odot$ and a shallower slope below that (see
Fig.~\ref {submassfunc}). A shallower slope results in a smaller value
of $\alpha$ and therefore $\gamma$.  Eq.~\ref{haneq} then predicts
that, compared to CDM, the radial number density profile of small mass
subhaloes should be suppressed less relative to the halo density
profile for subhaloes. This explains why the two lowest subhalo mass
bins in Fig.~\ref{subradialprofile} exhibit steeper radial density
profiles than the two highest mass bins.

An alternative way to examine the spatial distribution of
substructures is to plot the fraction of mass within a given radius
that is contained in substructures. This is shown in
Fig.~\ref{fracmass} for different ranges of host halo mass in \cococ~
and \cocow~. The radial distributions have roughly the same shape in
the two cases but the subhalo mass fractions are systematically lower
in \cocow~ than in {\sc coco-cold}. In both cases, the substructure
mass fractions are higher in more massive host haloes, particularly in
the inner regions. For example, for host haloes of mass
$M_{50}^{\mathrm{Host}}= (1-3) \times 10^{13}\,h^{-1}\,M_\odot$
resolved substructures in \cocow~ contain about $10\%$ of the halo
mass within $r = r_{50}$, but only about $4\%$ for host haloes of mass
$M_{50}^{\mathrm{Host}}=(1-3) \times 10^{11}\,h^{-1}\,M_\odot$.  For
reference, haloes (and subhaloes) contain $48\%$ of the total mass in
the simulation in \cocow~ and $56\%$ in \cococ~. In CDM simulations
these fractions depend on resolution, but not so in \cocow~ where the
cutoff in the power spectrum is resolved.

\subsection{Internal structure}
\label{internal}

The density profiles of WDM haloes and subhaloes are cuspy and well
described by the NFW \citep{Navarro1997} form
\citep{Lovell2012,Schneider2012}. However, the later formation times
of WDM haloes of mass near the cutoff scale, compared to their CDM
counterparts of the same mass, causes them to be less concentrated. In
\cite{Bose2016} we characterised the density and mass profiles of
haloes in \cocow~ over a range of halo masses and obtained the
concentration-mass relation for WDM haloes (see also
\citealt{Ludlow2016}).  In summary, the density profiles of the
largest haloes in \cocow~ (roughly two orders of magnitude above
$M_{\mathrm{hm}}$) are indistinguishable from their matched haloes in
\cococ~, but the profiles of haloes of mass $M_{200} < 7 \times
10^{10}\,h^{-1}\,M_\odot$ have systematically lower concentrations. In
contrast with the power-law concentration-mass relation in CDM, the
relation in WDM turns over at below $\sim 10^{10}\,h^{-1}\,M_\odot$.

Calculating the concentration of subhaloes from their density profiles
is not straightforward because the mass of a subhalo and therefore its
``edge'' are ambiguous. As \cite{Springel2008} showed, the size
calculated by the {\sc subfind} algorithm (that is the radius of the
saddle point in the density profile) coincides with the `tidal'
radius. Defining the concentration of the subhalo using this radius is
not particularly useful because its value varies along the orbit.
A more useful measure of subhalo concentration is the ratio
$V_{\mathrm{max}}/ r_{\mathrm{max}}$.  In both WDM and CDM, the
relation between $V_{\mathrm{max}}$ and $r_{\mathrm{max}}$ has a lower
normalisation for subhaloes than for ``field haloes'' because of tidal
stripping.

The fractional change in $V_{\mathrm{max}}$ between the moment of
infall and the present day is shown in Fig.~\ref{vmaxstrip} for
subhaloes (within $r_{50}$) of the most massive haloes in \cococ~ and
\cocow~, as a function of the present day maximum circular velocity,
$V_{\mathrm{max}}^{z=0}$ \citep[see also][]{diemand2007,pena2008}. The
largest subhaloes, with $V_{\mathrm{max}}^{z=0} \geq
50~\mathrm{kms}^{-1}$, experience a reduction in $V_{\mathrm{max}}$ by
a factor of $1.25-1.30$ after infall in both \cococ~ and \cocow~.
Subhaloes of lower mass show significant differences between the two
simulations.  For example, at $V_{\mathrm{max}}^{z=0} =
20~\mathrm{kms}^{-1}$, \cocow~ subhaloes have experienced a reduction
in $V_{\mathrm{max}}$ by a factor of $\sim 1.35$ since infall,
compared to $\sim 1.25$ for \cococ~ subhaloes.

The $r_{\mathrm{max}}- V_{\mathrm{max}}$ relations in \cococ~ and
\cocow~ are shown in Fig.~\ref{vrmax}. For large subhaloes the two are
very similar but the relations begin to diverge at values of
$V_{\mathrm{\max}}$ below of a few tens of kilometres per second,
depending on the mass of the host halo. In this regime, haloes of a
given $V_{\mathrm{\max}}$ have larger $r_{\mathrm{max}}$ in {\sc
  coco-warm} than in \cococ~ and are therefore less concentrated.  In
both \cococ~ and \cocow~ subhaloes are more concentrated than field
haloes, as a result of tidal stripping, but the difference between
field haloes and subhaloes is larger in \cocow~ than in \cococ~.  This
reflects the greater tidal stripping experienced by \cocow~ subhaloes,
which have lower concentrations when they fall into the host halo.  As
a result, the concentrations of subhaloes in \cocow~ increase more
than those in \cococ~ after infall.  Overall, however, \cocow~
subhaloes of a given mass (or \vmax~) still have lower concentrations
than \cococ~ subhaloes. As noted in \cite{Hellwing2016}, the
importance of tidal stripping depends weakly on host halo mass: at a
given $V_{\mathrm{max}}$, the reduction in $r_{\mathrm{max}}$ between
field haloes and subhaloes is slightly larger for larger host halo
masses

\begin{figure}
\centering
\includegraphics[width=\columnwidth]{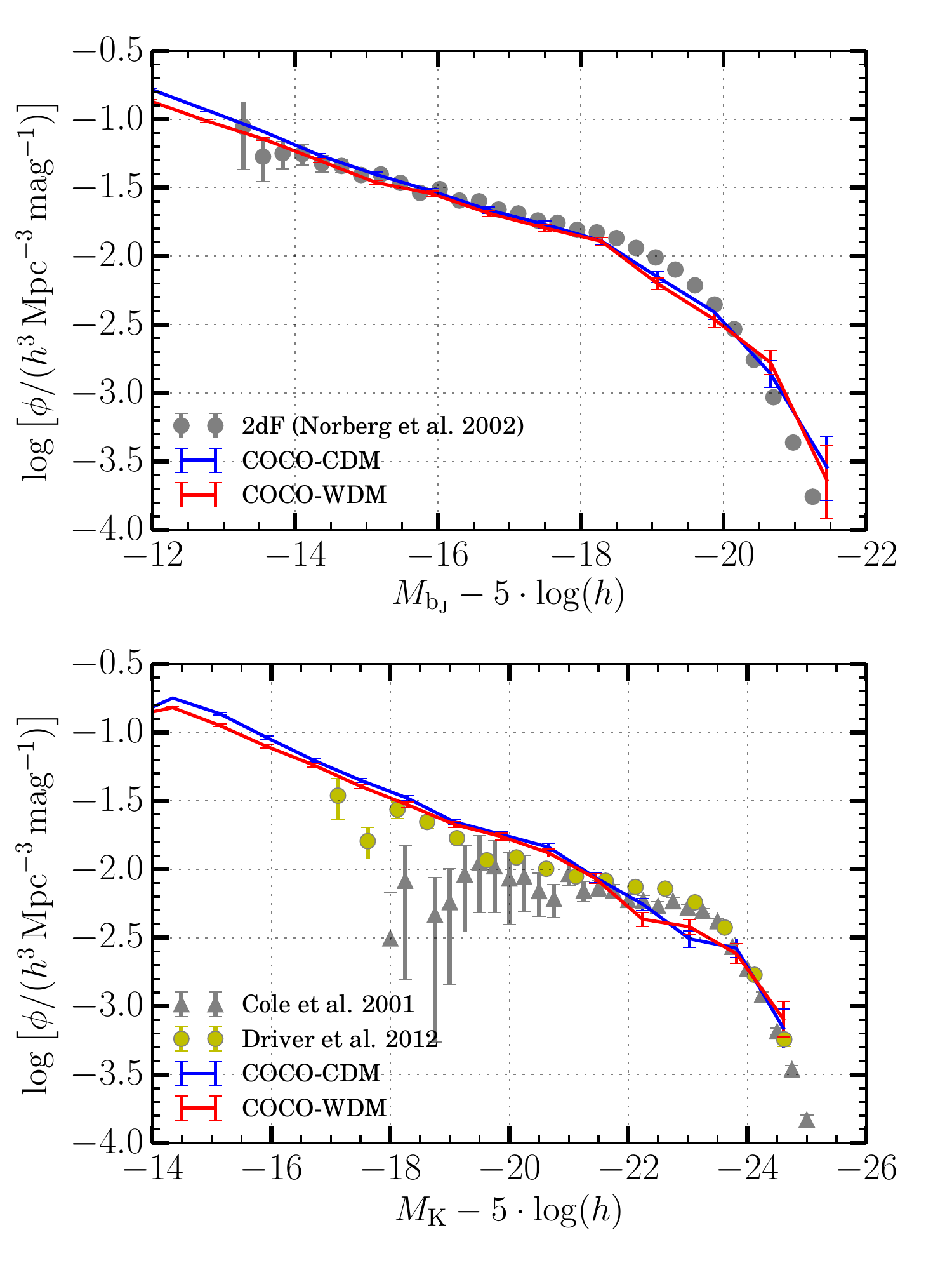}
\caption{The $z = 0$ $b_J$- (upper panel) and $K$-band (lower panel)
  luminosity functions from \textsc{galform} applied to halo merger
  trees constructed from the \cococ~ (blue) and \cocow~ (red)
  simulations (see text for details). The symbols represent
  observational data from \citealt{Norberg2002}, \citealt{Cole2001}
  and \citealt{Driver2012}.}
\label{lumfn}
\end{figure}

\begin{figure*}
\centering
\includegraphics[width=\textwidth]{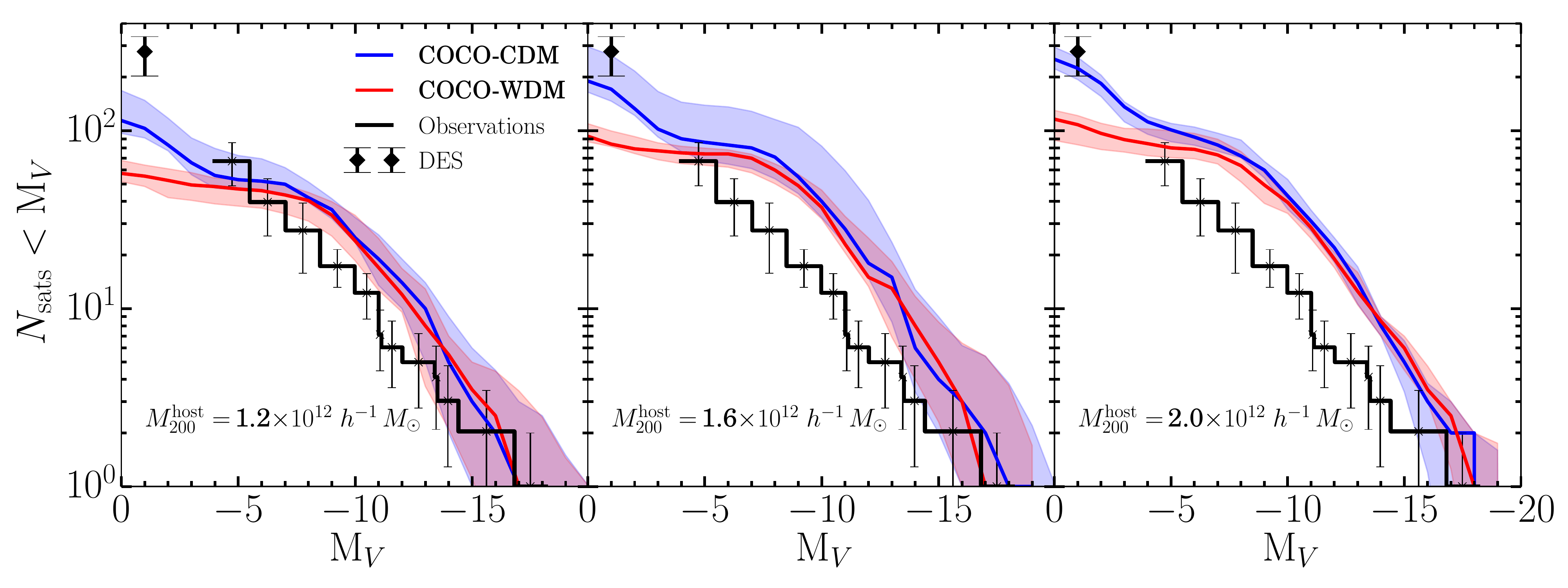}
\caption{The cumulative $V$-band luminosity function of satellites
  within $280\,\mathrm{kpc}$ of the centre of Milky Way-like haloes in
  \cococ~ (blue) and \cocow~ (red). Each panel shows the average
  luminosity function for host haloes in three bins of mass, $M_{200}
  =1-3\times 10^{12}\,h^{-1}\,M_\odot$, $1.5-1.7\times
  10^{12}\,h^{-1}\,M_\odot$ and $1.8-2.1\times
  10^{12}\,h^{-1}\,M_\odot$. The values quoted in the legend are the
  medians in each bin. The shaded regions indicate the 5th and 95th
  percentiles. The black step function shows the data for the Milky
  Way. For $\rm{M}_{\rm{V}} \geq -11$, the data has been corrected for
  incompleteness and sky coverage by \citealt{Koposov2008}.  For
  $\rm{M}_{\rm{V}} < -11$, the histogram shows the direct
  observational data from \citealt{McConnachie2012}. The black diamond
  is an extrapolation of the luminosity function to $M_{\rm V} \sim
  -1$ after including the ultra-faint dwarf satellites recently
  discovered by DES (\citealt{Jethwa2016}).}
\label{mwsat}
\end{figure*}

\section{Galaxy Formation with warm dark matter} 
\label{galaxygen}

Our analysis so far has been restricted to the dark matter properties
of a 3.3 keV thermal relic or, equivalently, a 7~keV sterile neutrino
with leptogenesis parameter, $L_6=8.66$, the ``coldest'' 7~keV sterile
neutrino compatible with the observed 3.5~keV X-ray line.  While
future gravitational lensing surveys may provide a direct way to
measure the mass function of dark matter substructures and thus
distinguish CDM from WDM \citep{Vegetti2009,RanLi2016}, it is worth
investigating whether CDM and WDM can be distinguished with current
observations.  At high redshift, the observed clumpiness of the
Lyman-$\alpha$ forest has been used to rule out WDM models with
thermally produced particles of mass $m_{\mathrm{WDM}} \leq 3.3$ kev
\citep{Viel2013}. As mentioned in Section~\ref{Intro}, constraints
obtained from the Lyman-$\alpha$ forest depend on assumptions for the
thermal history of the IGM.

To compare the models with other astronomical data we need to populate
the dark matter subhaloes with galaxies. This can be done in three
ways. One is to use empirical prescriptions such as ``abundance
matching'' \citep[see e.g.][]{Reed2015} but \cite{Sawala2015} have
shown that this technique breaks down for halo masses $<
10^{10}\,h^{-1}\,M_\odot$ -- precisely the scale of interest in
WDM. The failure of abundance matching in this regime is due to the
physics of reionisation, which inhibits the formation of stars in low
mass haloes after the epoch of reionisation, and to the effects of
supernovae feedback. A second technique are hydrodynamical simulations
but these are computationally expensive and, to date, only limited WDM
cosmological simulations have been carried out
(e.g. \citealt{Herpich2014,Carucci2015,GonzalezSamaniego2016}). The
third approach, the one we use here, is semi-analytical modelling of
galaxy formation, a flexible and powerful technique that requires only
modest computational resources.  We apply the Durham semi-analytic
model, \textsc{galform}, to halo merger trees in \cococ~ and
\cocow~. This model includes detailed treatments of gas cooling, star
formation, metal production, galaxy mergers and instabilities, black
hole growth and feedback from energy released by stellar evolution and
AGN. This model was previously used by \cite{Kennedy2014} to set a
lower limit to the mass of thermally produced WDM particles.

Details of the modelling in \textsc{galform} may be found in the
papers presenting the original formulation of the model
\citep{Cole2000} and its latest version \citep{Lacey2015}.  Here we
use this latest model for both \cococ~ and \cocow~ without any
modification\footnote{\cite{Kennedy2014} found that a small
  modification to one of the supernovae feedback parameters was
  required for their WDM models to produce acceptable $b_J$ and
  $K$-band luminosity functions at $z = 0$. The particle mass in the
  model we are considering here, 3.3~keV, is sufficiently large that
  not even this minor modification is required.}.

\subsection{Field and satellite luminosity functions}
\label{lumfuns}

The galaxy luminosity functions in the $b_J$ and $K$-bands in \cococ~
(see also \citealt{Guo2015}) and \cocow~ are compared with
observational data in Fig.~\ref{lumfn}.  The parameters controlling
supernova feedback in \textsc{galform} are calibrated to reproduce the
observed luminosity functions at $z=0$ in these bands. The two models
predict essentially identical luminosity functions except at faint
magnitudes where there are slightly fewer galaxies in WDM, as a result
of the lower abundance of small mass haloes in this model. At the
faintest magnitudes plotted the difference is only about 25\%, smaller
than the observational error bars. Due to the small volume of the
\textsc{coco} high resolution region, there are only a few bright
galaxies in the simulations, as reflected in the large Poisson errors
bars at the brightest magnitudes.

Fainter galaxies than those plotted in Fig.~\ref{lumfn} are only
detectable in the nearby Universe, particularly in the Local Group.
Only a few tens of satellites have been discovered orbiting the haloes
of the Milky Way and Andromeda. This number is much smaller than the
number of small subhaloes seen in CDM simulations of galactic haloes
and this observation has often been used to motivate WDM models. In
fact, it has been shown, using a variety of modelling techniques, that
most of these small subhaloes are not able to make a visible galaxy
either because their gas is heated by reionisation or expelled
altogether by supernovae explosions. The earliest explicit
demonstration of this simple physics was provided by the semi-analytic
models of \cite{Bullock2000} and \cite{Benson2002} and the latest by
the {\sc apostle} hydrodynamic simulations of \cite{Sawala2016}.

In fact, WDM models can be constrained by the observed number of faint
satellites because if the particle mass is too small not enough
subhaloes would form to account even for the observed number of
satellites in the Milky Way (which may be underestimated because of
incompleteness in current surveys). \cite{Kennedy2014} used this
argument to set constraints on the allowed masses of thermally
produced WDM particles. These constraints depend on the assumed mass
of the Milky Way halo because the number of subhaloes scales with the
mass of the parent halo (as seen, for example, in
Fig.~\ref{parentmsub} above). \cite{Kennedy2014} find that {\em all}
thermal WDM particle masses are ruled out (at 95\% confidence) if the
halo of the Milky Way has a mass smaller than $7.7 \times
10^{11}\,h^{-1}\, M_\odot$, while if the mass of the Galactic halo is
greater than $1.3 \times 10^{12}\,h^{-1}\, M_\odot$ only WDM particle
masses larger than 2~keV are allowed.

We perform a similar analysis here. Fig.~\ref{mwsat} shows the
cumulative number of satellites as a function of $V$-band magnitude,
$\mathrm{M}_V$, in \cococ~ and \cocow~ for three bins of host halo
mass, with median values of $1.2 \times 10^{12}$, $1.6 \times 10^{12}$
and $2.0\times 10^{12}\,h^{-1}\,M_\odot$. The luminosity function of
satellites in the Milky Way, shown by the black solid lines in the
figure, include the 11 classical satellites. For $\rm{M}_{\rm{V}} <
-11$, the data has been obtained from the direct observations of
\cite{McConnachie2012}. The abundance of ultra-faint satellites found
in the SDSS has been corrected for incompleteness and partial sky
coverage by \cite{Koposov2008}. The faint objects recently discovered
by DES \citep{Bechtol2015,DES2015} are represented by the black
diamond following the analysis of \citep{Jethwa2016} who find that of
the 14 newly-detected satellites, 12 have $> 50\%$ probability of
having been brought in as satellites of the LMC (at 95\%
confidence). \cite{Jethwa2016} extrapolate the detected population to
estimate that the Milky Way should have $\sim 180$ satellites within
300 kpc, in addition to $70_{-40}^{+30}$ Magellanic satellites in the
$V$-band magnitude range $-7 < M_{V} < -1$ (68\% confidence). All
observational error bars in Fig.~\ref{mwsat} are Poisson errors, with
volume corrections made where appropriate. In order to match the
observational selection, only satellites within $300\,\mathrm{kpc}$ of
the central galaxy are included.

The satellite luminosity functions are very similar in \cococ~ and
\cocow~. Only at magnitudes fainter than $\rm{M}_{\rm{V}}~\simeq~-4$
does the number of satellites in \cocow~ begin to drop below the
number in \cococ~.  The models agree with the data so long as the
Milky Way halo mass is $M_{200}^{\mathrm{host}}~\lsim~1.2~\times~
10^{12}\,h^{-1}\,M_\odot$. For
$M_{200}^{\mathrm{host}}~\sim~1.6~\times~ 10^{12}\,h^{-1}\,M_\odot$,
both \cococ~ and \cocow~ significantly overpredict the number of
satellites even at relatively bright magnitudes,~
$\rm{M}_{\rm{V}}~\sim~-10$, where the known sample is unlikely to be
significantly incomplete. There is a significant difference in the
abundance of satellites with magnitude $\rm{M}_{\rm{V}} \sim -1$, the
regime where DES has just begun to uncover ultra-faint dwarf
galaxies. These new data could potentially be used to set strong
constraints on the mass of the WDM particle. It must be borne in mind
that the exact location of this (extrapolated) DES data point depends
on the DES selection function, detection efficiency, and assumptions
made about isotropy in the distribution of Milky Way
satellites. Furthermore, although we have used a well-tested,
state-of-the-art model of galaxy formation, these conclusions depend
on assumptions in the model, particularly on the treatment of
reionisation and supernovae feedback \citep{Hou2016}.

\begin{figure*}
\centering
\includegraphics[width=0.9\textwidth]{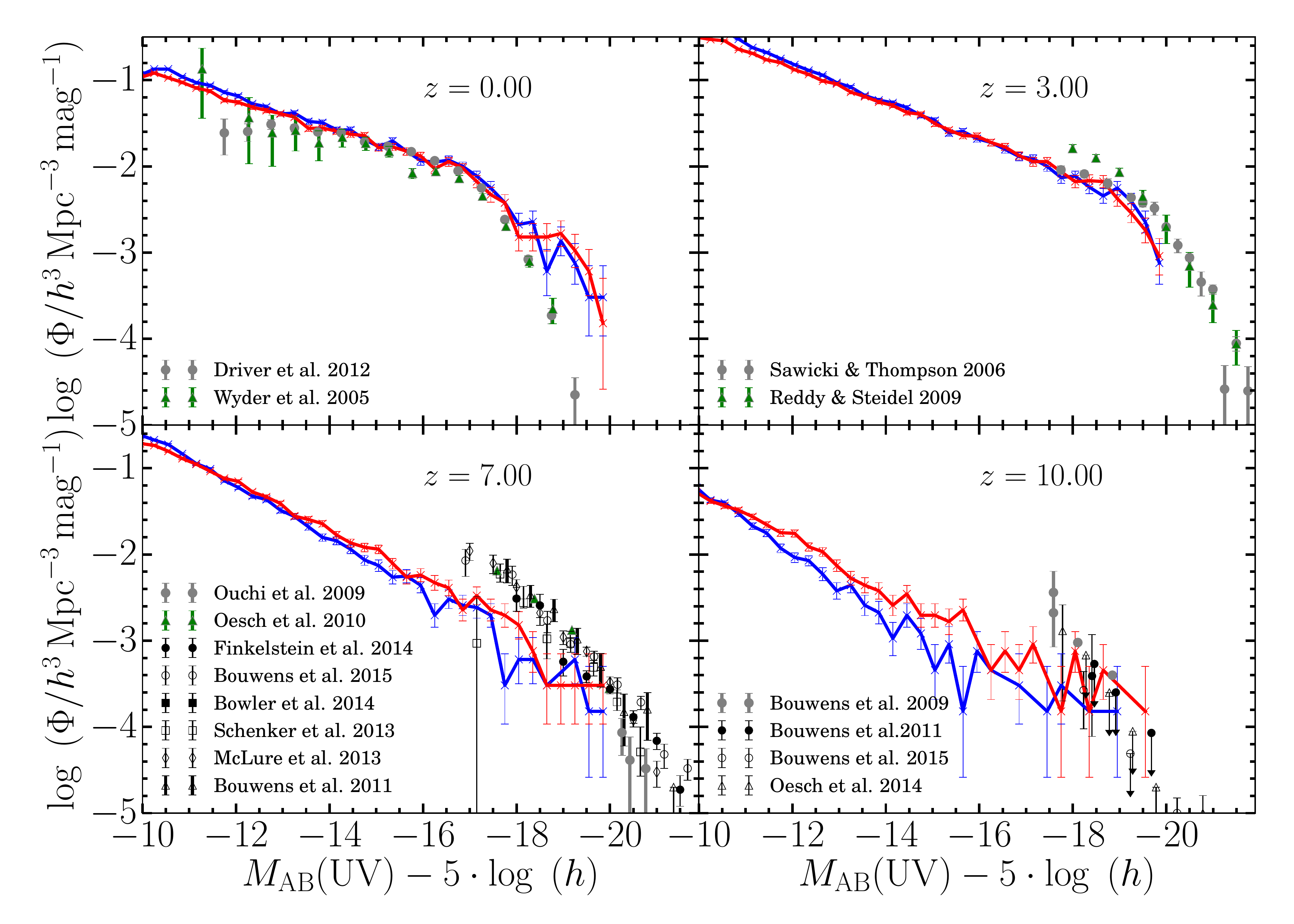}
\caption{The evolution of the UV luminosity function of all galaxies
  (centrals and satellites) for $z=0,3,7,10$. The red lines represents
  \cocow~ and the blue \cococ~, with Poisson errors plotted. The
  colour symbols with errorbars show observational data taken from
  \citealt{Driver2012}, \citealt{Wyder2005}, \citealt{Sawicki2006},
  \citealt{Reddy2009}, \citealt{Ouchi2009}, \citealt{Oesch2010},
  \citealt{Bouwens2009}, \citealt{Bouwens2011b,Bouwens2011},
  \citealt{Schenker2013}, \citealt{McLure2013},
  \citealt{Finkelstein2014}, \citealt{Bowler2014}, \citealt{Oesch2014}
  and \citealt{Bouwens2015}.}
\label{satuvlf}
\end{figure*}

\subsection{Evolution of the UV luminosity function}
\label{uvlf}

The evolution of luminosity function in the rest-frame UV traces the
star formation history in the Universe. Although still rather scarce
and uncertain, data now exist out to redshift $z \sim 10$. Since the
formation of structure begins later in WDM models than in CDM we might
na{\"i}vely expect to find fewer star-forming galaxies at high
redshift in \cocow~ than in \cococ~. The actual predictions are shown
in Fig.~\ref{satuvlf}, which reveals that, in fact, the result is
exactly the opposite: at $z>5$, the UV luminosity function has a
higher amplitude in \cocow~ than in \cococ~. The reason for this is
that, in CDM, supernovae-driven winds limit the reservoir of cold,
potentially star-forming, gas in low-mass galaxies at early times. The
brightest UV galaxies at high redshift tend to be starbursts triggered
by mergers of these relatively gas poor galaxies \citep{Lacey2015}. By
contrast in WDM, the first galaxies that collapse are more massive
than their CDM counterparts and more gas rich, thus producing brighter
starbursts when they merge. This makes the formation of bright
galaxies at high redshift more efficient in WDM than in CDM.

Although both \cococ~ and \cocow~ somewhat underpredict current
observations at $z>7$, the data have large statistical, and
potentially systematic errors since these objects are rare and current
surveys cover relatively small volumes. If anything, \cocow~ is closer
to the data than \cococ~. This result is broadly consistent with those
of \cite{Dayal2015} who used a simpler model of galaxy formation to
derive the UV luminosity function in WDM models.  The existence of a
population of star-forming galaxies in \cocow~ at $z>8$ has the
additional benefit that enough ionising photons are produced at early
times to reionise the universe by $z\simeq 8$, as required by the
optical depth to reionisation inferred from Planck \citep{Planck2013}.
Reionisation in WDM models is discussed in detail by \cite{Bose2016b}.

\begin{figure*}
\centering
\includegraphics[width=\textwidth]{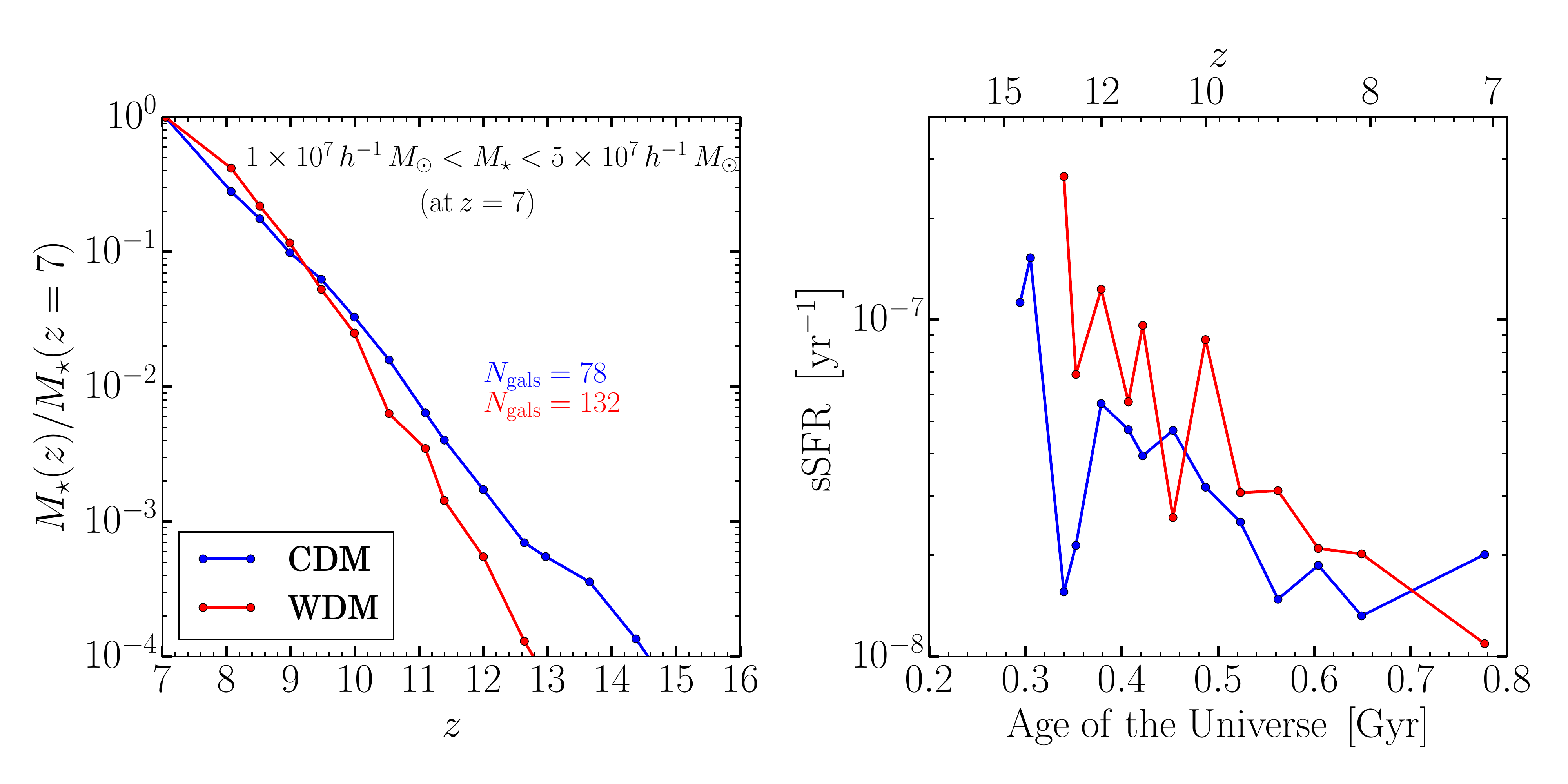}
\caption{Left panel: the average stellar mass growth of all galaxies
  with mass $1 \times 10^7\,h^{-1}\,M_\odot < M_\star < 5 \times
  10^7\,h^{-1}\,M_\odot$ in \cococ~ (blue) and \cocow~ (red). The mass
  as a function of redshift, $M_\star (z)$, is normalised to the final
  stellar mass at $z=7$. The number of galaxies averaged over in each
  simulation is indicated in the plot with the corresponding colour.
  Right panel: the specific star formation history as a function of
  the age of the Universe.  The galaxies averaged over are the same as
  in the left panel.}
\label{stellarmasshist}
\end{figure*}

Fig.~\ref{stellarmasshist} helps visualise the counter-intuitive
result just described. In the left panel we plot, as a function of
redshift, the stellar mass growth, $M_\star (z)$, averaged over all
galaxies with $1~
\times~10^7~h^{-1}~M_\odot~<~M_\star~<~5~\times~10^7~h^{-1}~M_\odot$
at $z=7$ in \cocow~ (red) and \cococ~ (blue). This range of stellar
mass corresponds to galaxies brighter than
$M_{\mathrm{AB}}~(\mathrm{UV})~\leq~-17$ in
Fig.~\ref{satuvlf}. $M_\star (z)$ is normalised to the stellar mass of
the galaxy at $z=7$, $M_\star (z=7)$. The stellar mass assembly in
\cocow~ is delayed relative to that in \cococ~ because the earliest
progenitors form later in \cocow~. For $12 > z > 8$, the build-up of
stellar mass is gradual in both \cococ~ and \cocow~, although the
slope of the mass growth is steeper in the latter i.e., more stellar
mass builds up per unit redshift in \cocow~ than in \cococ~.  This is
supported by the right panel of Fig.~\ref{stellarmasshist}, which
shows the evolution of the specific star formation rate (sSFR) of
these galaxies. \cocow~ galaxies exhibit systematically higher sSFRs
than \cococ~ up to $z=8$. This is in consistent with our earlier
suggestion that \cocow~ galaxies are formed out of more gas-rich
progenitors.  Mergers of these gas-rich progenitors allows galaxies in
\cocow~ to ``catch-up'' with those in \cococ~ after their delayed
start of star formation.

At $z\leq 3$ the UV luminosity functions in \cococ~ and \cocow~ are
indistinguishable even down to magnitudes as faint as
$M_{\mathrm{AB}} (\mathrm{UV}) \approx -10$. These galaxies form in
haloes of mass $\sim 10^{10}\,h^{-1}\,M_\odot$, the scale at which the
subhalo mass functions in \cocow~ just begin to diverge from those in
\cococ~ (see Fig.~\ref{submassfunc}). At even fainter magnitudes
($M_{\mathrm{AB}} (\mathrm{UV}) \geq -7$, not shown), the luminosity
function for \cocow~ is strongly suppressed relative to \cococ~ but
these magnitudes are far below the detection limits of even the JWST.

We have checked that the results in this section are not sensitive to
the specific version of the \textsc{galform} model used. The result in
Fig.~\ref{satuvlf} holds for the \cite{GonzalezPerez2014} model, with
and without the assumption of gradual ram-pressure stripping of hot
gas in satellite galaxies \citep{Font2008}, as well as for the
\cite{Hou2016} model in which supernova feedback is much weaker than
in our standard model at high-$z$ and becomes progressively stronger
at lower redshift. The simpler model by \cite{Dayal2015} is forced to
match the observed UV luminosity function at high-$z$ and cannot, by
construction, exhibit any differences between WDM and CDM.

\subsection{Other galactic observables}

In addition to the galaxy properties just discussed, we have explored
a number of others, such as colour and metallicity distributions;
sizes; the Tully-Fisher relation; and spatial clustering.  We do not
find any significant, potentially observable differences between
\cococ~ and \cocow~. This conclusion reinforces the point that, apart
from the details discussed in Section~\ref{lumfuns} and~\ref{uvlf},
galaxy formation is very similar in CDM and in a 7 keV sterile
neutrino (or a 3.3~keV thermal WDM) model.

\section{Summary and Conclusions}
\label{conclusion}
 
Using the {\it Copernicus Complexio} (\textsc{coco}) high resolution
dark matter simulations \citep{Hellwing2016}, we have carried out a
thorough investigation of the small-scale differences between CDM and
a model with the same phases but with a cutoff in the initial power
spectrum of fluctuations that can be interpreted either as that of the
``coldest'' sterile neutrino model compatible with the recently
detected 3.5~keV X-ray line or as a 3.3~keV thermal particle model.

The subhalo mass functions in the two models (\cococ~ and \cocow~) are
identical at high masses but the number density of \cocow~ subhaloes
begins to fall below that of \cococ~ subhaloes at $\sim 5 \times
10^9\,h^{-1}\,M_\odot$ and is very strongly suppressed below $\sim 2.5
\times 10^8\,h^{-1}\,M_\odot$, the half-mode mass in the initial power
spectrum, When the number counts are expressed in units of parent halo
properties such as $M_{\mathrm{sub}}/M_{200}$ and
$V_{\mathrm{max}}/V_{200}$, we find that the subhalo mass and
$V_{\mathrm{max}}$ functions in \cococ~ follow a nearly universal
profile with little dependence on host halo mass, confirming earlier
results \citep{Moore1999,Kravtsov2004,Zheng2005,
  Springel2008,Weinberg2008,Wang2012,Cautun2014}. This self-similar
behaviour does not occur in \cocow~.

The normalised radial distribution of subhaloes in both models is
independent of the mass of the subhaloes. In the case of \cocow~ this
behaviour extends to the smallest subhaloes in the simulation, with
$M_{\mathrm{sub}}\simeq 10^8\, h^{-1}\,M_\odot$, although there is a
slight steepening of their profile in the very central parts of the
halo. Our findings extend the results from the \textsc{aquarius} and
\textsc{phoenix} simulations \citep{Springel2008,Gao2012} and lend
support to the model proposed by \cite{Han2016} in which the mass
invariance of the radial distribution results from the effects of
tidal stripping. The radial density profiles are well approximated by
either the NFW or Einasto forms.

Subhaloes in both \cococ~ and \cocow~ are cuspy and follow the NFW
form. Small-mass WDM haloes, in general, are less concentrated than
CDM haloes of the same mass reflecting their later formation
epoch. For WDM subhaloes with $V_{\mathrm{max}}^{z=0}
\leq50~\mathrm{kms}^{-1}$, the difference is exacerbated because their
lower concentrations make them more prone to tidal stripping after
they are accreted into the host halo.

In order to check if the two models can be distinguished with current
observations, we populated the haloes with model galaxies whose
properties were calculated using the Durham semi-analytic galaxy
formation model, \textsc{galform}. We used the latest version of
\textsc{galform} \citep{Lacey2015} without needing to adjust any model
parameters for \cocow~. The \cococ~ and \cocow~ $b_J$ and $K$-band
luminosity functions at $z=0$ are very similar, except at the faintest
end where there are slightly fewer dwarfs in \cocow~; both models give
a good match to the observations. The same is true at the fainter
magnitudes represented by the satellites of the Milky Way: both models
agree with current data provided the mass of the Milky Way halo is
less than $M_{200} =1.2 \times 10^{12}\,h^{-1}\,M_\odot$. The two
models could be distinguished if the satellite luminosity function
faintwards of $M_V\sim -3$ or $-4$ could be measured reliably because
\cocow~ predicts about half the number of satellites as \cococ~ at
these luminosities.

The only other significant difference that we have found between
\cococ~ and \cocow~ is in the UV luminosity function at $z > 7$ where
there are more UV-bright galaxies in \cocow~ than in \cococ~. The
qualitative difference between the UV luminosity functions in \cocow~
and \cococ~ is not strongly affected by the treatment of baryon
physics in the \textsc{galform} semi-analytic model. This difference,
however (a factor of $\sim 2$ at $z>8$), cannot be detected with
current data. None of the other galaxy properties we examined: colour
and metallicity distributions, scaling relations, spatial clustering,
etc. differ in the two models in the regime where these properties can
be studied observationally.

In summary, the ``coldest'' sterile neutrino model compatible with the
identification of the recently detected 3.5~keV X-ray line as
resulting from the decay of these particles cannot, at present, be
distinguished from a CDM model by observations of galaxies, ranging
from the satellites of the Milky Way to the brightest starbursts at
$z=10$. The two models are drastically different in their dark matter
properties on subgalactic scales where the sterile neutrino model
predicts orders of magnitude fewer subhaloes of mass $M\lsim
10^8\,h^{-1}\, M_\odot$ than produced in CDM. These small masses are,
in principle, accessible to gravitational lensing
\citep{Vegetti2009,RanLi2016}, and it is to be hoped that future
surveys will be able conclusively to rule out one or the other or both
of these models.

 \section*{Acknowledgements}

We are grateful to Cedric Lacey for comments and help with {\sc
  galform}. This work was supported by: the European Research Council
grants GA 267291 (Cosmiway) and 646702 (CosTesGrav), the Science and
Technology Facilities Council [grant number ST/F001166/1,
  ST/I00162X/1, ST/K501979/1] and the Polish National Science Center
under contract \#UMO-2012/07/D/ST9/02785.  This work used the DiRAC
Data Centric system at Durham University, operated by the Institute
for Computational Cosmology on behalf of the STFC DiRAC HPC Facility
(\url{www.dirac.ac.uk}). This equipment was funded by BIS National
E-infrastructure capital grant ST/K00042X/1, STFC capital grant
ST/H008519/1, and STFC DiRAC Operations grant ST/K003267/1 and Durham
University. DiRAC is part of the National E-Infrastructure. Some
simulations used here were computed at the Interdiciplinary Centre for
Mathematical \& Computational Modelling at University of Warsaw with
the support of the HPC Infrastructure for Grand Challenges of Science
and Engineering Project, co-financed by the European Regional
Development Fund under the Innovative Economy Operational
Programme. This work is also part of the D-ITP consortium, a program
of the Netherlands Organisation for Scientific Research (NWO) that is
funded by the Dutch Ministry of Education, Culture and Science (OCW).

\bibliographystyle{mnras}
\bibliography{references.bib}{}

\label{lastpage}

\end{document}